\documentclass[iop]{emulateapj}

\usepackage{graphicx}
\usepackage{amsmath}
\usepackage{amssymb}
\usepackage{latexsym}
\usepackage{epsfig}

\newcommand{\half}{\frac{1}{2}}

\newcommand{\xv}{\mathbf{x}}
\newcommand{\nv}{\mathbf{n}}
\newcommand{\tv}{\mathbf{t}}
\newcommand{\vomega}{{\boldsymbol\omega}}
\newcommand{\rvir}{$R_\mathrm{vir}$}

\newcommand{\vvec}{\boldsymbol{v}}

\title{Turbulence in Galaxy Clusters}
\shorttitle{}
\shortauthors{F. Miniati}

\begin{document}
\title{The Matryoshka Run: Eulerian Refinement Strategy to Study
  Statistics of Turbulence in Virialized Cosmic Structures}

\author{Francesco Miniati}
\affil{Physics Department, Wolfgang-Pauli-Strasse 27,
ETH-Z\"urich, CH-8093, Z\"urich, Switzerland; fm@phys.ethz.ch}

\begin{abstract}
  We study the statistical properties of turbulence driven by
  structure formation in a massive merging galaxy cluster at redshift
  z=0. The development of turbulence is ensured as the largest eddy
  turnover time is much shorter than the Hubble time independent of
  mass and redshift. We achieve a large dynamic range of spatial
  scales through a novel Eulerian refinement strategy where the
  cluster volume is refined with progressively finer uniform nested
  grids during gravitational collapse. This provides an unprecedented
  resolution of 7.3 h$^{-1}$ kpc across the virial volume.  The
  probability density functions of various velocity derived quantities
  exhibit the features characteristic of fully developed compressible
  turbulence observed in dedicated periodic-box simulations. Shocks
  generate only 60\% of the total vorticity within \rvir/3 region and 40\%
  beyond that. We compute second and third order, longitudinal and
  transverse, structure functions for both solenoidal and
  compressional components, in the cluster core, virial region and
  beyond. The structure functions exhibit a well defined
  inertial range of turbulent cascade. The injection scale is
  comparable to the virial radius but increases towards the
  outskirts. Within \rvir/3, the spectral slope of the solenoidal
  component is close to Kolmogorov's, but for the compressional
  component is substantially steeper and close to Burgers'; the flow
  is mostly solenoidal and statistically rigorously consistent with
  fully developed, homogeneous and isotropic turbulence. Small scale
  anisotropy appears due to numerical artifact. Towards the virial
  region, the flow becomes increasingly compressional, the structure
  functions flatter and modest genuine anisotropy appear particularly
  close to the injection scale. In comparison, mesh
  adaptivity based on Lagrangian refinement and the same finest resolution,
  leads to lack of turbulent power on small scale and excess thereof on large scales,
  with the discrepancy growing towards the outer cluster regions, while producing
  unreliable density weighted structure functions throughout.
\end{abstract}

\section{Introduction} \label{intro:sec} 
\subsection{Modeling Turbulence}

Astrophysical flows are characterized by large Reynolds numbers,
$\mathrm{Re}\equiv \frac{uL}{\nu_k},$ where $L$ and $u$ are the
characteristic spatial scale and velocity of the fluid motions,
respectively, and $\nu_k$ the kinematic viscosity. A large Re implies
the existence of a wide range of spatial scales where the fluid
interactions are dominated by the nonlinear term in Euler's equation,
while momentum diffusion remains negligible. This leads to the cascade
of fluid motions predominantly from larger to smaller scales, thus
generating turbulence~\citep{LandauLifshitz6}.  Turbulence affects the
qualitative behavior of a fluid by adding new features to it. To
mention a few, turbulent random motions alter the stability of a
self-gravitating fluid by contributing an additional effective
pressure~\citep{Chandrasekhar51,Schmidt13}.  In a compressible
isothermal flow, they also generate density peaks which may turn
gravitationally
unstable~\citep{ElmegreenScalo04,Schmidt10,Konstandin12}. Turbulence
modifies the transport coefficient of a fluid, e.g. the diffusion of
scalar quantities such as temperature, entropy and metals
~\citep[e.g.][]{Zeldovich90}.  Turbulence strongly amplifies magnetic
field~\citep[e.g.][]{zeruso83,biskamp93,ChoYoo12,Federrath11,Beresnyak12}.
It is argued that turbulence changes the fundamental workings of
magnetic reconnection in astrophysical
plasma~\citep{LazarianVishniac99,Lazarian12}.  For these and many
other reasons the role of turbulence is investigated in several
subjects of physics and astrophysics, ranging from the earth's upper
atmosphere, the sun, the interstellar medium and molecular clouds of
galaxies, and galaxy clusters.

An important effort in the study of turbulence concerns the
determination of the structure of the fluid quantities as produced by
the turbulent cascade. Numericists pursue this effort using large
(magneto) hydrodynamical simulations, with periodic boundary
conditions, in which the turbulence is driven on as large a scale as
allowed by the computational box, in order to maximize the dynamic
range of spatial scales. The turbulence is studied under variety of
conditions including, amongst others, incompressible and compressible
flows, steady state and decaying regime, solenoidal and/or
compressible forcing~\citep{Federrath08,Federrath13}, 
isothermal or adiabatic-law equation of state.
This is partly motivated by the variety of conditions that apply to
different astrophysical systems.  The aim is to understand the
fundamental properties of the turbulence, including the scaling
relations that apply to velocity and other thermodynamic variables
and, related to this, the structures responsible for energy
dissipation. One may then wish to use this understanding to properly
model actual astrophysical objects in accord with their physical
conditions.  Models of driven turbulence in periodic boxes reduce to a
minimum the complications associated with boundary effects and,
provided numerical convergence, they allow for a clean interpretation
of the results.

It is also important, however, to have the ability to model the
turbulent flows directly in the actual astrophysical context and
environment.  This allows to produce more realistic conditions under
which the turbulence develops. In particular, the forcing terms may
not necessarily operate on a narrow range of scales, may include a
combination of solenoidal and compressible terms, may very well be
time dependent. The flow itself may be self-gravitating and non
homogeneous. So, while the periodic-box studies remain instrumental
for the analysis of the results, there is value in modeling the
turbulence under the proper astrophysical conditions.  This, of
course, has remained of limited applicability because of the
considerable numerical requirements, as turbulence does not develop
unless the Re of the flow is large enough.

With the foregoing objectives in mind, in this paper we attempt to
study the statistical properties of the turbulence in the intracluster
medium (ICM) of a massive galaxy cluster (GC). In particular we use a high
resolution hydrodynamical simulation of structure formation to compute
probability distributions function and structure functions of various
velocity fields to establish the genuine turbulent character of the
fluid motions inside the GC volume. Here we focus on a purely
hydrodynamic model (the magnetic field is present but completely
negligible for the dynamics) and leave the case of saturated
magneto-hydrodynamical turbulence to future work. 

Applying mesh adaptivity based on Lagrangian refinement criterion
produces high resolution in only a small fraction of the volume.  This
results in well resolved collapsed structures but the effective Re of
the flow remains poor. So while large eddies appear, turbulence cannot
develop. To obviate this and achieve a sufficiently large range of
well resolved spatial scales across the GC volume we employ instead a
novel Eulerian refinement approach, providing uniform high spatial
resolution throughout the GC volume.  Uniform resolution also avoids
biases in the statistics and allows us to carry out the analysis as
closely as possible to the periodic-box cases.

There are several previous numerical studies of turbulence in the ICM
but with very different focus (see below).  The work
of~\cite{Vazza09,Vazza11} is closest to ours in terms of purposes
although there are still differences in terms of peak resolution and
dynamic range, resolution strategy and analysis.  Analogous efforts in
other fields include the study of three-dimensional convective
turbulence in the stratified atmosphere of
stars~\citep{Porter00,Viallet13} and the magneto-rotational instability
in the magnetized accretion disks~\citep{Bodo11,Parkin13}.


\subsection{Turbulence in Galaxy Clusters}

Galaxy clusters (GC), with mass up to a few $\times 10^{15}$M$_\odot$,
are the largest virialized systems, sitting at the top of the
hierarchical chain of cosmic structure.  They grow hierarchically,
through mergers with other already collapsed systems, and also through
smooth accretion of matter that is still far from virialization.
Massive GC are still dynamically young and in the process of
formation.  As a result, they have yet to reach fully relaxed
dynamical conditions.  Most of the baryonic mass is in a state of
fully ionized hot plasma namely the ICM, with
number density and temperature in the range, $n_e\sim
10^{-2}-10^{-5}$cm$^{-3}$ and, $T\sim 10^7-10^8$K, respectively, from
the core to the outskirt region.

From the hydrodynamical point of view both the merger process and
smooth accretion play important roles (see also Sec.~\ref{tim:sec}).
Mergers set up large scale motions in the ICM, which eventually
dissipate through turbulent cascade or shocks of various strength.
Smooth accretion occurring through filaments typically penetrates
inside GC down to a fraction of the virial radius depending on the
filament size, generating large scale shear flows and shocks at the
filament-ICM interface.  Smooth accretion from voids is hypersonic and
produces very high Mach number shocks with curvature radii of a few
Mpc. As shown later (Sec.~\ref{tdm:sec}) these shocks are
characterized by highly irregular surface and only partially dissipate
the kinetic energy of the accreting gas. Owing to their curvature,
they generate large scale vorticity, thus acting as an additional
sources of turbulence.

The development of turbulence relies on large Re.  This depends on the
mean free path of thermal particles in the ICM which is somewhat subject
to debate. Magnetic fields, however, are commonly observed in the
ICM~\citep{Clarke01,Clarke04,FerettiHollitt04}. So it is assumed here
that the mean free path is determined by the Larmor radius in the
direction perpendicular to the magnetic field, and by microscopic
(e.g. mirror and firehose) plasma instabilities along the mean
magnetic field lines~\citep{Parker58,Ginzburg79,Schekochihin05},
although the simulation resolution never reaches scales below the
Coulomb mean free path in the ICM. The important thing is that,
irrespective of the specific viscosity mechanism, using $10^3$ km
s$^{-1}$ for the typical flow velocity and a Mpc for the
characteristic spatial scales (Sec.~\ref{tim:sec}) leads to large
enough Re ($\gtrsim 10^3$) for the onset of turbulent motions
throughout the GC volume.

The properties of turbulence in the GC are important for a variety of
reasons. Support against gravity from turbulence as well as bulk
motions introduces biases in the GC mass estimates based on the
assumption of hydrostatic
equilibrium~\citep{Faltenbacher05,Dolag05b,Lau09,Nagai07,Biffi11,Valdarnini11}, with
important consequence for the use of GC as cosmological probes.
Characterizing the turbulence allows us to determine the expected
level equipartition between magnetic and kinetic energy in the ICM and
shed light on the origin of magnetic fields
there~\citep{Subramanian06,Iapichino08}.  Cluster turbulence affects
the transport of relativistic particles both in momentum and
configuration space, and is studied also to understand the origin of
cluster diffuse radio sources~\citep{Paul11,Hallman11,Vazza09,Vazza11}.

Observations of turbulence in the ICM are, however, extremely
challenging.  To first attempt to determine the presence of
turbulent motions on a range of scales in the ICM is due
to~\citet{Schuecker04}. These authors used X-ray XMM-Newton
observations of the nearby massive Coma GC, to measure the pressure
fluctuations induced by turbulent motions in an incompressible
fluid~\citep{Kolmogorov41a,Kolmogorov41c,Oboukhov41a}.
The presence and spectral properties of turbulence have also been
probed via the analysis of the structure of rotation measure (RM)
maps, of extended radio sources typically embedded in the
ICM~\citep{EnsslinVogt03,KucharEnsslin11}, leading to results
consistent with a Kolmogorov spectrum.

In fact, turbulent motions imprint a number of features on the ion
emission lines due to Doppler effect, such as profile broadening and
centroid shifting~\citep[e.g.][]{Sunyaev03}. For heavy ions these
effects dominate over thermal broadening ($\propto 1/m_{ion}^{1/2}$),
opening a new observational window on the properties of turbulent
motions~\citep{InogamovSunyaev03,Zhuravleva11,Zhuravleva12}.  This
science will become possible with the advent of high precision X-ray
spectrometers, such as Astro-H\footnote{http://astro-h.isas.jax a.jp}
and Athena\footnote{http://sci.esa.int/ixo}, which will have energy
resolution of a few eV, allowing to measure gas motions of order 100
km/s in massive clusters.  Currently, Doppler broadening of iron
emission lines is employed to probe turbulent motions in cool cores,
where thermal motions are particularly
low~\citep{Churazov08,Sanders10,Sanders11}.  Alternatively, the
presence of turbulent gas motions in the ICM (of Perseus cluster) has
been inferred from the lack of resonant scattering effects for the
He-like iron emission line at 6.7 keV~\citep{Churazov04}.

The rest of this paper is organized as follows.  Full details of the
numerical calculation, including the code, the initial condition, the
cosmological model and the refinement strategy, are described in
detail in Sec.~\ref{Numerics:sec}.  The main results, in particular
the analysis of the simulations in terms of two-dimensional plots,
probability density functions, and various velocity structure
functions, convergence analysis and comparison with
Adaptive-Mesh-Refinement, are presented in Sec.~\ref{t_res:sec}. The
discussion in Sec.~\ref{t_disc:sec} concludes the paper.

\section{Numerics} \label{Numerics:sec}

\subsection{Code, Cosmology and Initial Conditions}\label{code:sec}

The simulation of structure formation is carried out with the
Adaptive-Mesh-Refinement cosmological code
\texttt{CHARM}~\citep{MiniatiColella07b}.  We use a directionally
unsplit variant of the Piecewise-Parabolic-Method for
hydrodynamics~\citep{Colella90}, a time centered modified symplectic
scheme for the collisionless dark matter and solve Poisson's equation
with a second order accurate discretization~\citep{MiniatiColella07b}.
In addition to hydrodynamics and self-gravity, the calculation evolves
a dynamically negligible magnetic field using the
constrained-transport algorithm for solenoidal MHD described
in~\cite{MiniatiMartin11}. Radiative cooling and heating of the
intergalactic medium through photoionization is neglected. While this
results in a lower gas temperature in voids and, in a correspondingly
higher Mach number for the outermost accretion shocks, it has no
consequences whatsoever for the generation of vorticity and turbulence
at shocks.

We assume a concordance $\Lambda$-CDM universe with normalized (in
units of the critical value) total mass density, $\Omega_m=0.2792$,
baryonic mass density, $\Omega_b=0.0462$, vacuum energy density,
$\Omega_\Lambda= 1- \Omega_m= 0.7208$, normalized Hubble constant
$h\equiv H_0/100$ km s$^{-1}$ Mpc$^{-1}$ = 0.701, spectral index of
primordial perturbation, $n_s=0.96$, and rms linear density
fluctuation within a sphere with a comoving radius of 8 $\,h^{-1}$
Mpc, $\sigma_8=0.817$~\citep{Komatsu09}.

The initial conditions are generated for a volume of comoving size
$L_{Box}=240\,h^{-1}$ Mpc, sufficiently large to accommodate
statistically significant sample of massive clusters at the turnover
of the mass function.  We use \texttt{grafic++}, the parallel version of
the \texttt{grafic2} package~\citep{Bertschinger01}, developed and made
publicly available by D. Potter and the power spectrum interpolation
suggested in~\cite{EisensteinHu98}.

We carry out a preliminary low resolution run using a uniform grid of
512$^3$ comoving cells, corresponding to a nominal spatial resolution
of 468.75$\,h^{-1}$ comoving kpc. We sample the distribution function
of the collisionless dark matter component with 512$^3$ particles with
mass $6.7\times 10^9\,h^{-1}$ M$_\odot$.  At redshift zero halos are
identified using our implementation of the HOP halo
finder~\citep{EisensteinHut98}, adopting the standard parameters
suggested in the original finder paper.  A massive GC, with a mass
around 10$^{15}$ M$_\odot$, is then selected for re-simulation at high
resolution.

Zoom-in initial conditions are then generated, using again the 
\texttt{grafic++} code.  The matter that ends up inside the simulated
massive GC is collected from a Lagrangian volume of about 20$\,h^{-1}$
comoving Mpc in radius. This volume is initialized at a spatial
resolution $\Delta x=$ 117.2$\,h^{-1}$ comoving kpc, typically
sufficient to resolve structures 100-1000 times smaller than the final
GC. To achieve this, we use two additional levels of refinement on top
of the base grid.  The refinement ratio for both levels is, $n_\mathrm{ref}^\ell\equiv \Delta x_{\ell}/\Delta x_{\ell+1}=2$, $\ell=0,1$.
Each refined level covers 1/8 of the volume of the next coarser level
with a uniform grid of 512$^3$ comoving cells.  The dark matter
distribution function is likewise represented by 512$^3$ particles on
each refined level, so that at the finest level where the initial
conditions are generated the particle mass is $1.0\times 10^8\,h^{-1}$
M$_\odot$.

\subsection{Eulerian Refinement Strategy}\label{app:sec}

\begin{figure*}[t] 
\centering
\includegraphics[width=0.6\textwidth,angle=90]{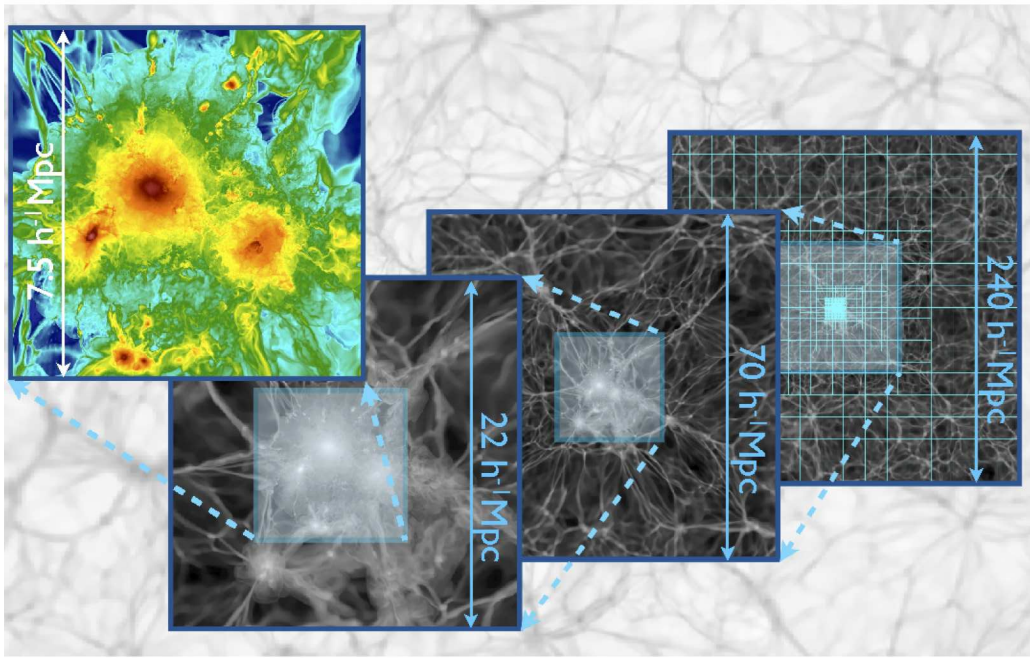}
\caption{
From left to right, panel sequence showing the two dimensional
slices of the baryonic gas density distribution in log scale, 
closing in on the massive GC at the center of the computational volume.
The plots correspond to cosmological redshift $z=0.45$.
In the leftmost panel, the full
hierarchy of five nested grids with progressively higher resolution is
also shown, superposed to the gas density distribution.
The translucent bluish-shaded quadrants in the first three 
panels show the region resolved with finer grids at the next level
of refinement. The rightmost panel shows the innermost and
highest resolution grid.
\label{t_f1:fig}}
\end{figure*}

An important objective of our calculation is to resolve with a certain
degree of accuracy the inertial range of the turbulent cascade
expected to develop in the ICM. The turbulent cascade is initiated at
the injection scales, so our aim is to model the turbulent velocity
field in a range of scales below injection such that it remains
unaffected by numerical viscosity.  According to dedicated studies of
compressible turbulence, this requires that the scales of interest be
resolved with about 32 resolution
elements~\citep{Porter92,Porter94,Porter02,Federrath11}.  The injection scale of
the turbulence, $L_{inj}$, is expected to be in the range 300-1000
$\,h^{-1}$kpc~\citep{NormanBryan99a,Schuecker04,Vazza09}, therefore we
aim for a mesh size on the finest grids of order of a several
$\,h^{-1}$kpc, so that turbulent motions above 100 $\,h^{-1}$kpc will
be resolved.  This requirement is easily achieved if one employs the
AMR technique the way it is typically applied in cosmological
simulations, i.e. using a criterion for mesh refinement
that is based on mass content of the volume element. We refer to 
this as Lagrangian flavor of AMR, as it aims at keeping the mass
resolution constant, like in particle based Lagrangian codes.

However, this method of refinement is clearly not satisfactory for our
purposes because small scales are biased toward high density regions
and turbulence will remain unresolved in most of the GC volume.
This was already recognized in~\cite[e.g.][]{Iapichino08,Paul11}, where
the mass threshold refinement criterion was replaced with one based on the
vorticity of the flow, and also in~\cite{Vazza09,Vazza11} where instead a
refinement threshold criterion based on discontinuities of the
velocity field was used.  As a result of these new refinement criteria, the GC
volume fraction covered by high resolution grids increased
dramatically.

However, applying finer resolution elements based on the local fluid
properties may not necessarily be the best approach.  In fact, the AMR
technique is of limited computational efficiency when the filling
factor is large, e.g. larger than a few 10\%, which is certainly the
case for fully developed turbulence~\citep{Kritsuk06}.  In addition,
the presence of coarse/fine grid interfaces introduce numerical error
compared to a uniform grids~\citep{BergerColella89}. This is not
desirable when one is interested in studying the statistical properties,
particularly high order ones, of velocity fluctuations associated to
turbulence, as it may introduce spurious contribution.

%
\begin{table}  [t]
\begin{center}
\begin{small}
\caption{Eulerian Refinement Strategy\label{run:tab}}
\begin{tabular}{cccccc}
\hline
\hline
{$\ell$} &
{T} &
{$L$} &
{$N_\ell$} &
{$n_\mathrm{ref}^\ell$} &
{$\Delta x_\ell$} \cr
& $(H_0^{-1})$ & ($h^{-1}$Mpc) &  & & ($h^{-1}$kpc) \cr
\hline
0 & 0 & 240 &  512 & 2  & 468.7 \cr
1 & 0 & 120 &  512 & 2  & 234.4 \cr
2 & 0 & 60  &  512 & 2  & 117.2 \cr
3 & 0.013 & 30  &  512 & 4  & 58.6 \cr
4 & 0.23 & 15  & 1024 & 2  & 14.6 \cr
5 & 0.48 & 7.5 & 1024 & -  & 7.3  \cr
\hline
\hline 
\end{tabular}
\end{small}
\end{center}
\end{table}
Therefore, we have adopted a different approach to achieve the desired
high spatial resolution in the GC volume.  Basically, we use a set of
nested and progressively finer grids that cover uniformly the volume
occupied by the simulated GC.  The grids are generated dynamically as
the Lagrangian volume of the simulated GC shrinks in size under the
pull of its self-gravity.  We refer to this as Eulerian refinement
strategy. Table~\ref{run:tab} summarizes the details of our zoom-in
simulation.  We use a total of 5 levels of refinement. For each level
(1), the table report the time of activation (2), the size on a side
of the cubic volume covered (3), the number of cells on a side of the
uniform grid (4), the refinement ratio with respect to the next finer
level (5), and the mesh size (6), respectively.  The additional
refinement levels (3-5) are employed flexibly as the GC assembles.
One can anticipate/delay their activation at the higher/lower cost of
a larger/smaller refined region.  Basically a new refinement level is
generated when it will be able to accommodate at least two-thirds of
the GC Lagrangian volume.  In addition, if required, the highest
density peaks can also be refined independently based on the
Lagrangian refinement criterion, although we did not find this
necessary.  Note that in most cases a refinement ratio
$n_\mathrm{ref}^\ell=2$ is used, except for level $\ell=3$, which uses
$n_\mathrm{ref}^3=4$, to achieve more efficiently the desired
resolution in the GC volume.
As shown in Table~\ref{run:tab}, with our Eulerian refinement strategy
we achieve very high, uniform spatial resolution across the
virial volume of the simulated GC.  This allows us to study in great
detail not only the cluster core region, but also the GC outskirts,
and determined how the property of the ICM turbulence vary across the
virial volume. 

Figure~\ref{t_f1:fig} shows, as an example, the hierarchy
of nested grids when the simulation run has reached approximately
redshift 0.4, when the GC is experiencing a major merger.
In the leftmost panel, a two dimensional slice of the baryonic gas
density distribution is shown in log scale, with superposed the five
nested grids of the hierarchy, with sizes listed in
Table~\ref{run:tab}.  The rightmost panel shows the innermost and
highest resolution grid.

The calculations required 450 base level time-steps.  This corresponds
to ca 8,700 and 11,000 steps on the fourth and fifth level of
refinement, respectively, for a total of 2$\times 10^{13}$ MHD solver
updates.  The latter is the most expensive part of the code and the
two finest levels is where most of the CPU cycles are used.
Both the calculation and the data analysis required running on several
thousand cores on a Cray XE6 at the Swiss National Supercomputing
Center for a total cost of several hundreds of thousand of CPU hours.

\section{Results} \label{t_res:sec} 
The analysis below is restricted to data at redshift $z=0$. 
The evolution of the presented
results as a function of redshift will be presented elsewhere.  We
note however, that the simulated GC suffered a major merger around
redshift 0.2, and has accumulated roughly half of its virial mass
since then, 20\% of which during the last $\half$Gyr.  So it is
currently in the process of strong dynamical relaxation.

\subsection{Characterization of the Galaxy Cluster}
\begin{table}  [t]
\begin{center}
\begin{tiny}
\caption{Simulated GC Properties\label{gc:tab}}
\begin{tabular}{cccccc}
\hline
\hline 
{M$_{vir}$} &
{R$_{vir}$} &
{R$_{200}$} &
{R$_{500}$} &
{T} &
{v$_\mathrm{rms}$} 
 \cr
($10^{15}\,$M$_\odot)$ & ($h^{-1}$Mpc) & ($h^{-1}$Mpc) & ($h^{-1}$Mpc) & (keV) & (km$\,s^{-1}$) \cr
\hline
1.27 & 1.95 & 1.5 & 1 & 5.2 & 1927 \cr
\hline
\hline 
\end{tabular}
\end{tiny}
\end{center}
\end{table}
We start with a description of the basic properties of
the simulated GC.  We run again our HOP halo finder to determine
its center.  We then determine various characteristic quantities
including the virial radius enclosing a mass over-density
$\Delta_c=178\, \Omega_m^{0.45}$~\citep{Eke01} with respect to the
critical density, $R_\mathrm{vir}=1.95\,h^{-1}$ Mpc, and the
corresponding enclosed mass, $M_\mathrm{vir}=1.27\times 10^{15}$
M$_\odot$.  Other characteristic radii corresponding to higher over
densities often used in the literature, include, $R_{200}\simeq
1.5\,h^{-1}$ Mpc, $R_{500}\simeq 1\,h^{-1}$ Mpc. The volume averaged
gas temperature is 5.2 keV and the gas volume averaged rms velocity is
1927 km s$^{-1}$. These values are summarized in Table~\ref{gc:tab}.

\subsubsection{Radial Profiles}
\begin{figure}[t]
\centering
\includegraphics[width=0.5\textwidth]{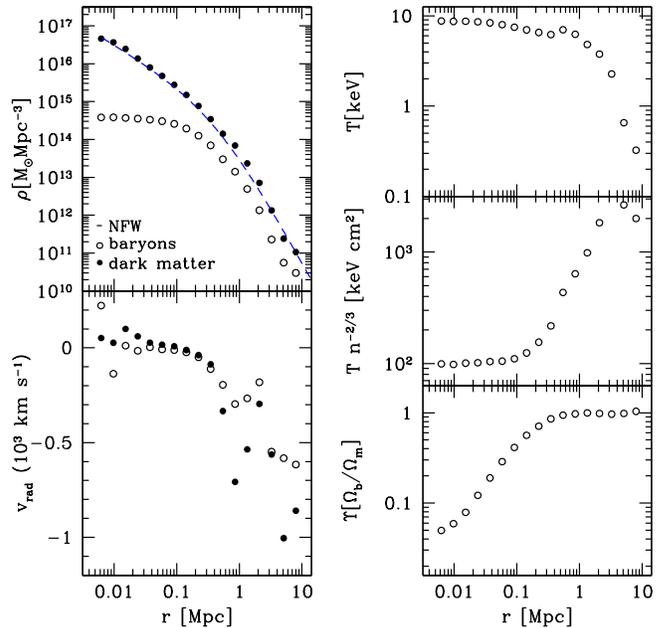}
\caption{\textit{Top Left:}
Density profile of the baryonic gas (open circles)
and dark matter (filled circles). The dash line correspond
to a NFW profile with concentration parameter $c=7.5$.
\textit{Bottom Left:} Profile of the radial velocity component
for the baryonic (open circles) and dark matter (filled circles).
The structure around 1 Mpc distance indicates the presence of 
a large infalling clump.
\textit{Top Right:} Gas temperature profile. The bump
around 1 Mpc radial distance again is due to the presence of
infalling clump.
\textit{Middle Right:} Volume averaged specific entropy radial profile.
\textit{Bottom Right:} Cumulative radial profile of the baryonic gas 
fraction, in units of the cosmic average value.
The plots correspond to cosmological redshift $z=0$.
\label{t_f2:fig}}
\end{figure}
\begin{figure*}
\centering
\includegraphics[width=0.9\textwidth,angle=-90]{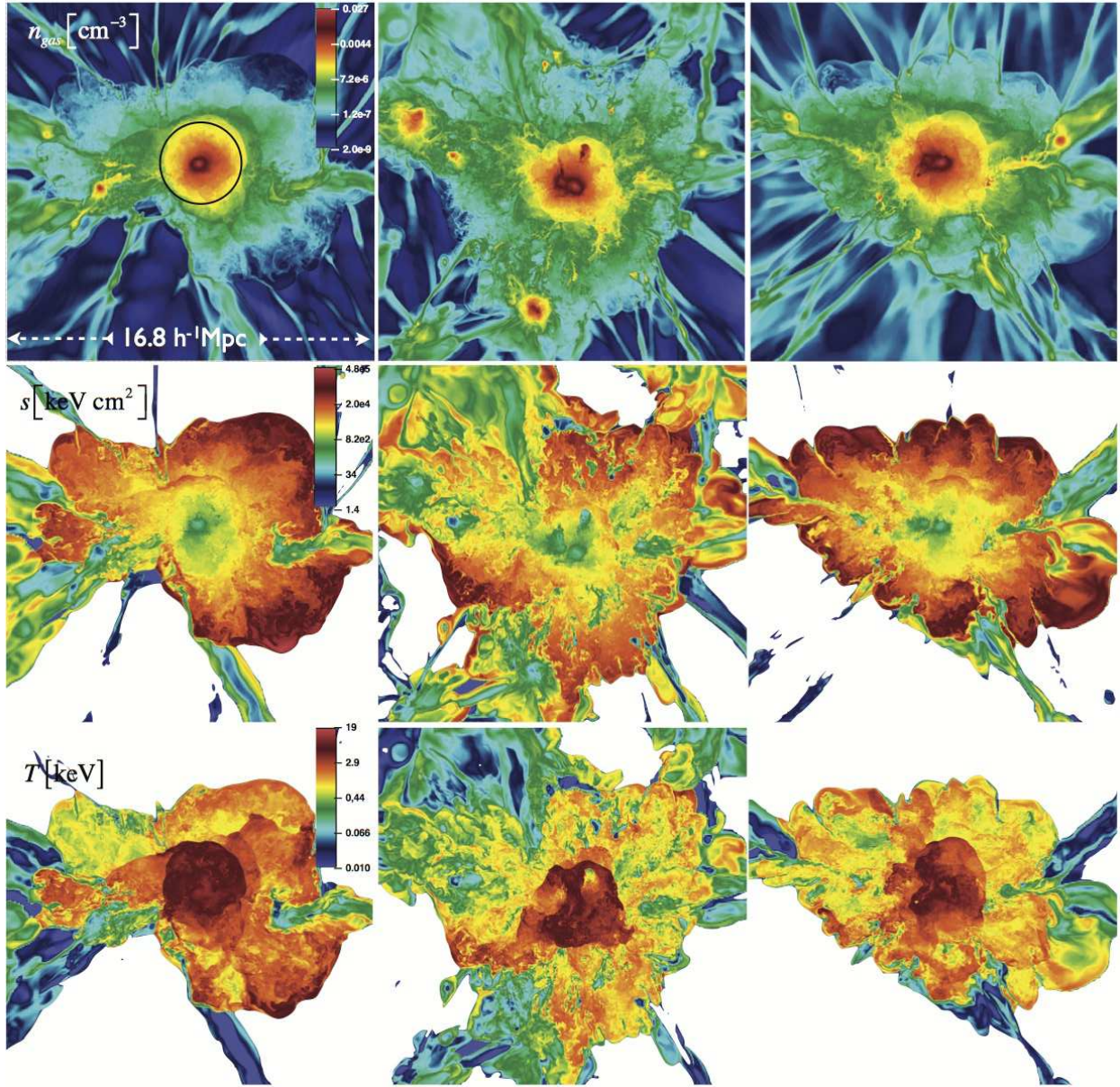}
\caption{Two dimensional slices of density (top)
gas specific entropy (second from top), and temperature (second from
bottom), for three different
planes passing through the GC center (left to right).
Each panel is 16.8$\,h^{-1}$Mpc on a side. 
The black circle in the top-left panel indicates the region 
enclosed within the virial radius.
Color bar in physical units for each quantity is shown at the
top-left corner of the leftmost panel.
\label{t_f3:fig}}
\end{figure*}
\begin{figure}
\centering
\includegraphics[width=0.45\textwidth]{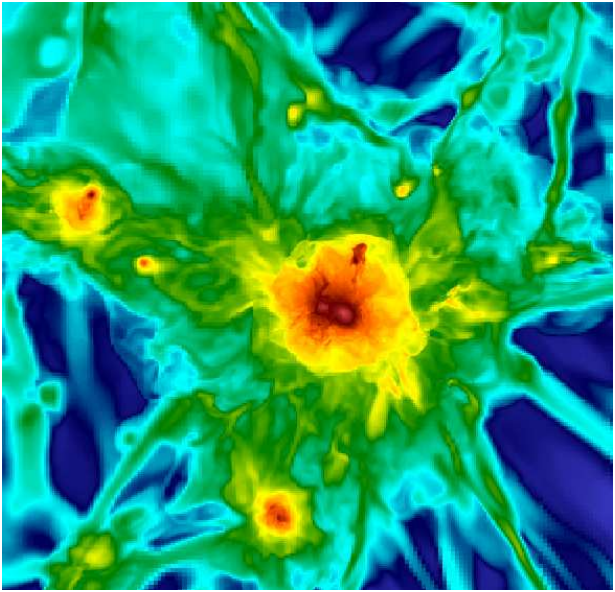}
\caption{Two dimensional slices of density, just as in
the central top panel of Fig.~\ref{t_f3:fig}, but from an
AMR simulation based on Lagrangian refinement criterion.
\label{t_f4:fig}}
\end{figure}
\begin{figure*}
\centering
\includegraphics[width=0.9\textwidth,angle=0]{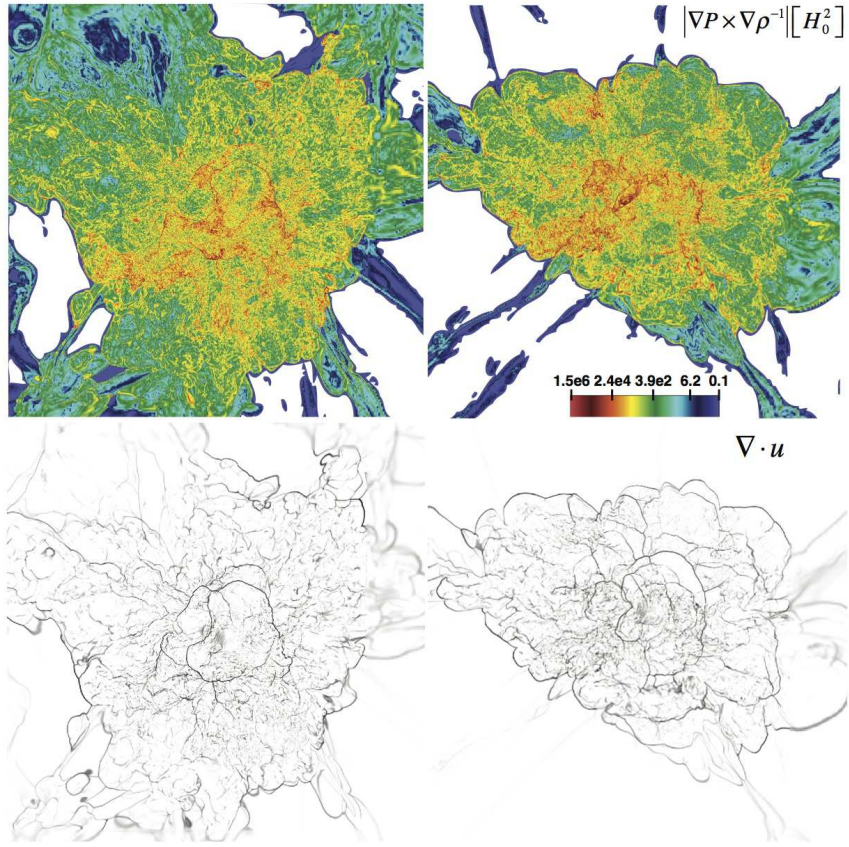}
\caption{Two dimensional slices of baroclinic term's magnitude (top)
and velocity divergence (bottom), for two different
planes passing through the GC center (left to right).
Each panel is 16.8$\,h^{-1}$Mpc on a side. 
Color bar for the baroclinic plot is shown in 
the top-right panel in physical units of $H_0^2$.
\label{t_f5:fig}}
\end{figure*}
\begin{figure*}
\centering
\includegraphics[width=0.9\textwidth,angle=0]{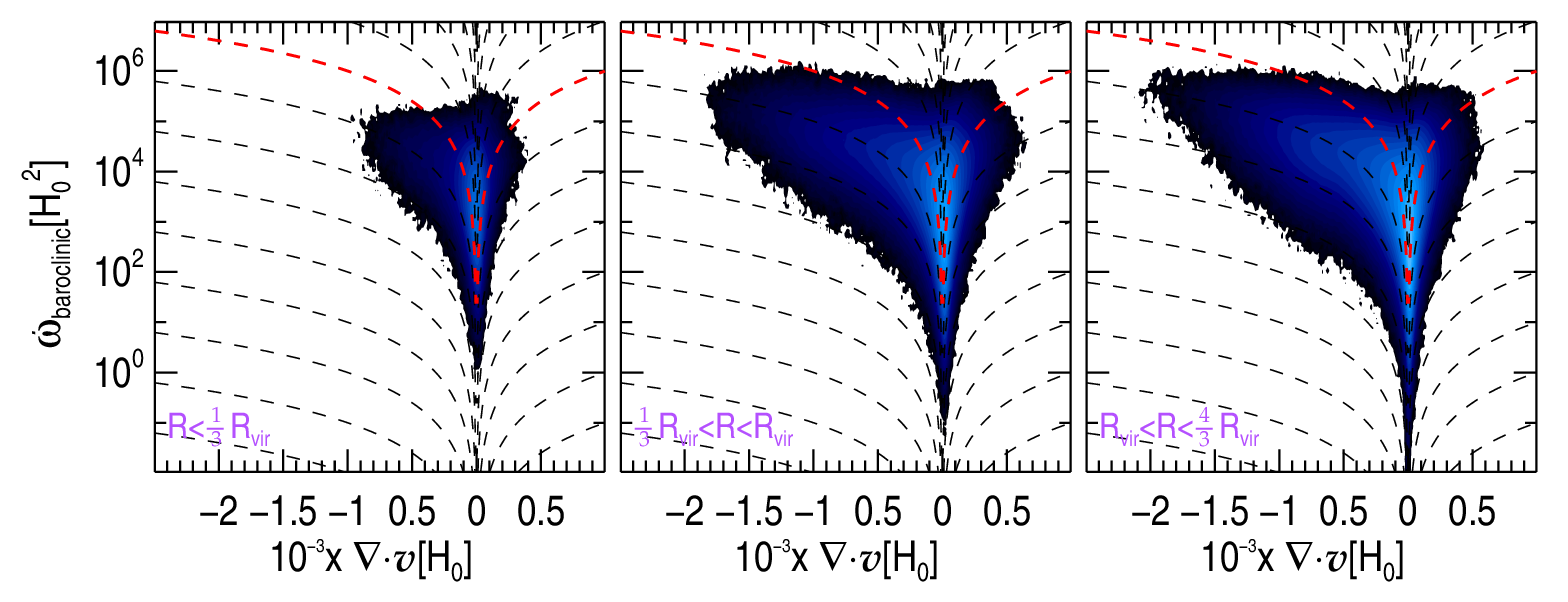}
\caption{Phase-space diagram of the baroclinic term's magnitude
($\dot\omega_{\rm baroclinic}$) in units of $H_0^2$ versus 
velocity divergence in units $10^{3}H_0$. The three panels correspond to the
core region (left), the off-core and virial region (center), and 
the off-virial region (right). The dash diagonal lines
are loci where $\dot\omega_{\rm baroclinic}=A(\nabla\cdot v)^2$, with $A$ 
given by the label in the leftmost panel. 
The red line corresponds to $A=1$. For lines below and above it,
the factor $A$ is progressively smaller and larger by a factor 10, respectively.
\label{t_f6:fig}}
\end{figure*}
In Figure~\ref{t_f2:fig} we show the radial profiles of additional
basic quantities. The top left panel illustrates the density profiles
for the baryonic gas (open circles) and dark matter component (filled
circles), respectively. The dash line correspond to a NFW profile with
concentration parameter $c=7.5$. This is shown for illustration only,
as no effort was made in fitting the precise value of this parameters,
and we found that $c=8$ also provides a viable choice to match the
numerical profile.  The plot shows the typical core distribution of
the baryonic gas (open circles), where the average gas density is
$n\simeq$ a few $\times 10^{-2}$cm$^{-3}$, versus the cuspy profile of
the dark matter (filled circles).  The bottom left panel shows the
profile of the radial velocity component, again open circles for
baryonic gas and filled circles for dark matter component,
respectively. One can see velocity structure both in the inner region
at radial distances around 10 kpc, and particularly a prominent
feature around 1 Mpc. These are due to the presence of large infalling
clumps. A feature around 1 Mpc, originating from the same event, is
also visible in the temperature profile (top right panel).  While
volume average gas temperature is 5 keV, the baryonic gas in the GC
core reaches hotter temperatures, around 9 keV, and stays relatively
hot even beyond $R_{vir}$, with $T\sim$ 1 keV.  The middle right panel
shows the radial average entropy distribution. The radial profile
rises steadily way beyond $R_{200}$ and bends slightly only beyond the
virial radius. It turns over only at radii a few times
$R_\mathrm{vir}$, roughly the location of the external accretion
shocks, beyond which the lay the intergalactic medium.  Finally, the
bottom right panel show the cumulative radial profile of the baryonic
gas in units of the cosmic average value. In this simulation, the
baryonic gas fraction reaches the cosmic average well within the
virial radius.

\subsubsection{Two-dimensional Maps}\label{tdm:sec}
Figure~\ref{t_f3:fig} shows two dimensional slices of density (top), gas
specific entropy (middle), and temperature (bottom), respectively, for 
three different
planes passing through the GC center (left to right).  The size of the
panel on a side is about 16.8$\,h^{-1}$ Mpc. The center of the GC is
roughly located at the density peak in the top panels.  The region
enclosed within virial radius, $R_{vir}$, is illustrated by the black
circle.  

The first point to notice is that, as illustrated by
Figure~\ref{t_f3:fig}, the virial volume of the GC is embedded in a
much larger volume filled with hot gas and enveloped by the external
accretion shocks at a few times the virial radius.  We refer to the
gas hot within the external accretion shocks, including the hot gas
beyond the virial radius, i.e. formally outside the GC volume, still
as the ICM.
The external accretion shocks, where the temperature jumps from sub eV
to keV temperatures, are found 2-3 $R_{vir}$ from the GC center. This
is much further out than the virial radius, as well as the radius
$R_{200}$ ($\sim 3/4$ R$_\mathrm{vir}$), which X-ray observations are now
starting to explore.

From the morphological perspective, the overall shape of the ICM in
Figure~\ref{t_f3:fig} appears very irregular.  The maps on the left
panel are strongly asymmetric, the one on the right elongated, the one
in the center very extended. The surface of the accretion shocks also
appear highly structured and irregular.  Large scale asymmetries are
revealed by maps of various thermodynamic quantities which appear
quite different on different planes.  In the following Section we will
describe the role of these large scale irregularities, including the
shock surface structures, in generating turbulence.  Perhaps the most
striking features is the tremendous amount of structure exhibited in
the maps, particularly in the GC outskirts. This is due to fluid
instabilities and turbulent motions. Their development is made
possible by the relatively high Re number allowed by the uniform high
numerical resolution across the GC volume.  For comparison, in
Figure~\ref{t_f4:fig} we show the same density map as shown in the
top central panel of Figure~\ref{t_f3:fig}, but obtained with a
simulation which is identical in all respects to the one being
presented here, e.g. in terms of initial conditions, cosmological
parameters, max spatial resolution etc., except that it employs
adaptive mesh refinement based on Lagrangian criterion. Specifically
grid cells are tagged for refinement when their mass content is eight
times as large as the mean comoving value.  Simple visual inspection
shows that while the two simulations are fully consistent with each
other in terms of gross flow features, in the Lagrangian AMR case
one does not see fine structure of the flow as fluid instabilities
hardly develop on small scales. A more quantitative analysis based
on velocity structure functions is presented in Section~\ref{amrcomp:sec}.

The small and large scale fluid structure in the maps of
Figure~\ref{t_f3:fig} is due to the fact that the system is
dynamically active, i.e. is still forming.
Specific entropy traces the thermal history of a fluid element due to
non-adiabatic processes. Lacking any endothermic and/or exothermic
processes, the only non-adiabatic processes here are those due to
shock heating.  The specific entropy maps allows us to recognize the
presence of clumps tracing merging substructures that are floating
around the GC potential, and filaments, in addition to the GC core.
All of them have formed relatively early.
The gas in the clumps and filaments, subject to various fluid
instabilities such as Rayleigh-Taylor, Kevin-Helmholtz, turbulent
drag, is stripped and deposited in the ICM. Mixing with the ICM is,
however, incomplete.  The motion of the substructures traced by low
entropy gas is particularly important in generating turbulence in the
ICM.
The temperature maps (second from bottom) show also large
inhomogeneities.  These are partly due to the presence of gas with
different thermal history as discussed above, which include the clumps
but also the large filament in the leftmost panel. Another important
source of temperature structure is the presence of large scale shocks,
extending across a large fraction of the GC volume.  These internal
shocks play an important role in the generation of turbulence, which
is discussed in the next section.

The properties of the turbulent flow across the GC vary with radial
distance. For this reason in the following we typically distinguish
four different radial regions in which we carry out the analysis.
These are defined as: \textit{core} with
$R<\frac{1}{3}R_\mathrm{vir}$, \textit{off-core} with
$\frac{1}{3}R_\mathrm{vir}<R<\frac{2}{3}R_\mathrm{vir}$,
\textit{virial} with $\frac{2}{3}R_\mathrm{vir}<R<R_\mathrm{vir}$,
and, \textit{off-virial} with
$R_\mathrm{vir}<R<\frac{4}{3}R_\mathrm{vir}$.

\subsection{Turbulence Injection Mechanisms}\label{tim:sec}

In view of the analysis presented below, it is useful to discuss
the various processes that contribute to the generation of the
turbulence in the ICM.
Turbulence is generated by shearing flows and the baroclinic term.
Shearing flows are unstable to turbulence when the Reynolds number is 
sufficiently large.
On the other hand 
the baroclinic term generates vorticity, $\vomega\equiv\nabla\times\vvec$, 
according to~\citep{LandauLifshitz6},
\begin{equation}\label{ome:eq}
\frac{\partial\vomega}{\partial t} =
\nabla\times(\vvec\times\vomega)+ \nabla P\times \nabla\frac{1}{\rho},
\end{equation}
which in turn excites turbulence.  The baroclinic term appears when
the fluid pressure depends explicitly on both density and temperature,
$p=p(\rho,T)$, i.e. the fluid is baroclinic.  Shocks are a strong
source of baroclinicity but, as shown below, not the only one.  Both
shear flows and the baroclinic term can be traced back to
\begin{enumerate}
\item tidal fields
\item merging substructure
\end{enumerate}
with gravity the ultimate source of energy.
\begin{figure*}[t]
\centering
\includegraphics[width=1\textwidth,angle=0]{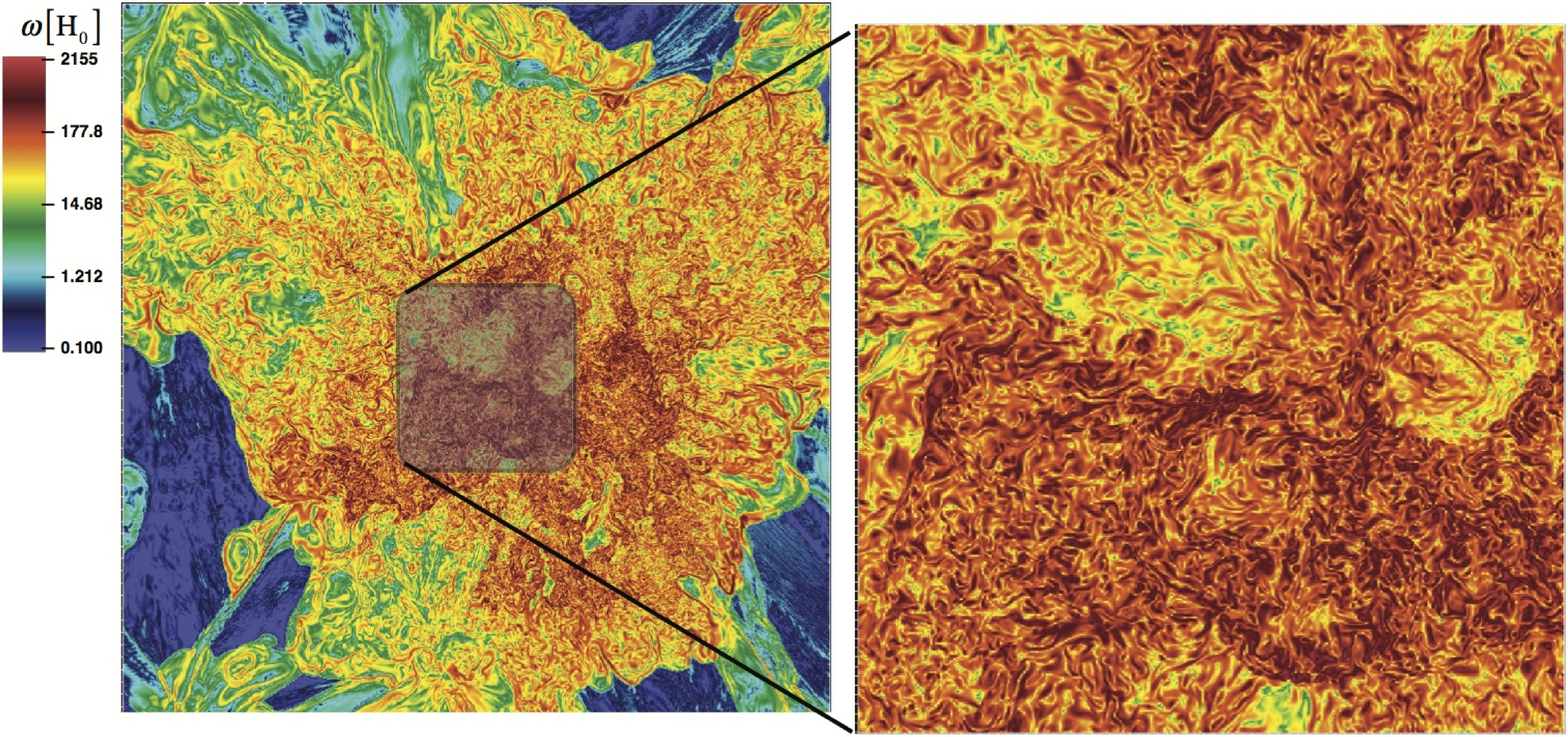}
\caption{\textit{Left:} Two dimensional slice of 
vorticity magnitude on a plane 
passing through the GC center. 
The plane orientation is as in the middle panels of Figure~\ref{t_f3:fig}.
\textit{Right:} Zoom-in of the shaded area in the left panel.
Vorticity is expressed in units of the Hubble's constant, with 
color-bar on the top-left of the figure.
\label{t_f7:fig}}
\end{figure*}
\begin{figure*}
\centering
\includegraphics[width=0.9\textwidth,angle=0]{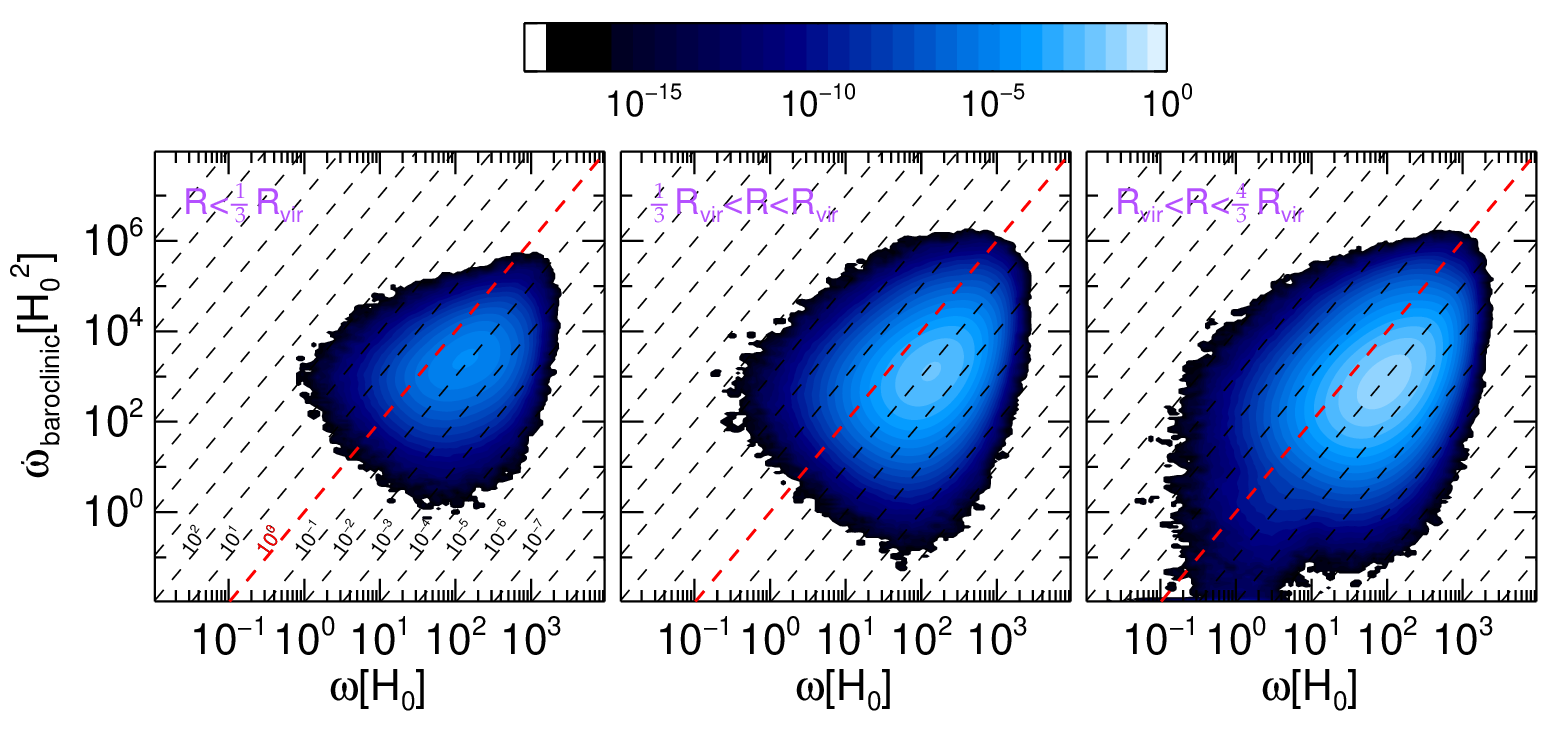}
\caption{Phase-space diagram of the baroclinic term's magnitude
($\dot\omega_{\rm baroclinic}$) in units of $H_0^2$ versus 
vorticity in units $H_0$. The three panels correspond to the
core region (left), the off-core and virial region (center), and 
the off-virial region (right). The dash diagonal lines
are loci where $\dot\omega_{\rm baroclinic}=A\omega^2$, with $A$ 
given by the label in the leftmost panel, and with 
the red line corresponding to $A=1$.
\label{t_f8:fig}}
\end{figure*}

Due to the presence of tidal fields in the volume around the GC, the
velocity field of the accretion flow is anisotropic, and the accretion
shocks acquire an irregular shape.  This is important for two reasons.
Firstly, as a result of the changing curvature of the shock surface,
vorticity is generated by the baroclinic term.  Secondly, the
accretion shocks have an \textit{oblique} character, i.e. they are
quite inefficient at dissipating the upstream flow kinetic
energy~\citep{Miniati00}. Downstream the accretion shocks residual
trans and supersonic motions persist, and the gas temperature remains
modest ($\sim$ 1 keV).  The interaction of these residual motions
generates turbulent
cascades, as well as large scale shocks internal to the ICM,
particularly in the region around and beyond the virial radius.  The
internal shocks in turn generate vorticity through the baroclinic
term, which is particularly strong at sites of shock-shock
collisions. These shocks, therefore, further contribute to the
generation of turbulence.

The top panels of Figure~\ref{t_f5:fig} shows two dimensional maps of
the baroclinic term for two different planes passing through the GC
center (the same planes in the central and right panels of
Fig.~\ref{t_f3:fig}).  The two bottom panels illustrate the position
of the shocks in the same planes through the negative divergence of
the velocity. The figure shows that the ICM is pervaded by a fine
network of predominantly weak shocks.
These shocks generate a baroclinic term all over the ICM, with
the strength of the latter depending on the strength of the former.
However, the baroclinic term appears more volume filling than
shocks. This reveals that shocks are not the only source of
baroclinicity.  This is better illustrated in Figure~\ref{t_f6:fig},
which presents the phase-space diagram of the baroclinic term versus
velocity divergence. The three panels correspond to the core (left),
off-core + virial (center), and off-virial (left) region,
respectively. Clearly the baroclinic term is active not only in
regions of negative velocity divergence (shocks), but also in regions
where the latter quantity is null or positive, i.e.  without
shocks. In fact, due to the hierarchical assembly of a GC, the ICM is
composed of mixtures of gas with different thermodynamic
histories. In particular, there coexist polytropic gas with different
entropy. This implies that the ICM deviates significantly from a
barotropic fluid, i.e.  is generally baroclinic, and generation of
vorticity does not necessarily require shocks.  In fact, in the
specific case illustrated here, shocks generate baroclinicity in 60\%
of the cases in the core region, but only 40\% everywhere else (out to
the off-virial region).  This may sound counter-intuitive, but is
related to the fact that while the gas in the outskirts regions is
more compressional, shocks there are not as pervasive as in the GC
inner regions (see Figure~\ref{t_f5:fig}).

The spatial scale characterizing the gradient of the velocity field
associated to the accretion and the residual post-shock flows can be
estimated of order the curvature radius of the external and internal
shocks, which in turn is of order the virial radius, i.e. $L\sim
R_\mathrm{vir}$. The velocity field can be estimated using the scaling
relation, yielding, $v_\mathrm{vir}=\sqrt{GM_\mathrm{vir}/R_\mathrm{vir}}$.
Therefore, the characteristic rate associated to the turbulent cascade is
the dynamical timescale of a virialized structure,
\begin{equation}\label{tr:eq}
\tau_\mathrm{turb}^{-1}(z) \simeq
\frac{v_\mathrm{vir}}{R_\mathrm{vir}}\sim \Delta_c^{\half} H(z)\gg H(z),
\end{equation}
and is fast compared to the cosmological timescale. Note that this statement
is true for collapsed halos regardless of redshift.

Merging halos and substructures floating through the GC potential,
also contribute to stirring up energetic motions in the ICM.  In
particular, turbulence is generated in the wakes of these halos as
they move through the ICM~\citep{Subramanian06}.  The injection scale is
around the halo ram pressure stripping radius.  The volume filling
factor is very sensitive to the Reynolds number, which determines the
length of the wake~\citep{Subramanian06}.  Only during major mergers,
i.e. mergers with structures of similar mass, is turbulence injected
on large scales, $L\lesssim$R$_\mathrm{vir}$, and with large volume
filling factor.  For most substructure, the injection scale is
considerably smaller, i.e. $L~\sim$ 100 kpc, for a sub-halo mass $m
\simeq 10^{13}$M$_\odot$. The filling factor is uncertain due to the
unknown viscosity of the ICM, but remains small compared to unity if
viscosity is provided by Coulomb collisions.  In addition to the turbulent wakes,
moving substructure generate bow shocks which are weak during core
passage, but steepen as they enter the low density regions
at and beyond the virial radius.  These shocks generate vorticity and,
hence, turbulence as discussed above.  The injection scale can be
estimated a few times the characteristic size of the halo, which is
given by the above ram pressure stripping radius, i.e. $L~\sim$ a few
$\times$ 100 kpc.

Because in general $L\lesssim$R$_\mathrm{vir}$, and the velocity involved
is of order $v_\mathrm{vir}$, the conditions expressed by
Eq.~(\ref{tr:eq}) remain valid, i.e. there is sufficient time for the
motions generated by mergers and moving substructures to generate a
turbulent cascade during a Hubble time at virtually any redshift.

Finally, the filaments through which the merging halos accrete onto
GC, provide another example of anisotropic accretion and source of
turbulence.  Filaments of different sizes penetrate through the GC
atmosphere, with the largest filaments reaching the core region.  The
strong shear flow that is present at the ICM-filament interface is
subject to Kelvin-Helmholtz instability.  Inspection of
Figure~\ref{t_f3:fig} shows that filaments inside the ICM are indeed
unstable, with the small sized ones being completely disrupted. The
instability extracts the free energy of the gas in the filaments and
feeds turbulence.  Additionally, large filaments typically terminate
with a shock, which is a source of additional turbulence as discussed
above.  Filaments come in different sizes, but the characteristic
length, $L$, scales for the injection of turbulence by the instability
and by the filament termination shocks are of the same order as
discussed above for the substructure.  Therefore, Eq.~(\ref{tr:eq})
applies as well in this case.

In conclusion, large scale motions appear to be generated on a range
of spatial scales by different processes. The largest scales reach up
to $R_\mathrm{vir}$, or a fraction thereof, and are contributed by tidal
fields and major mergers. Smaller structure stirs fluid motions on
accordingly smaller scales.  In any case, the characteristic time
$L/v$ is in general much smaller than the Hubble time.

\subsection{Signatures of Fully Developed Turbulence}\label{fdt:sec}
\begin{figure}[t]
\centering
\includegraphics[width=0.5\textwidth]{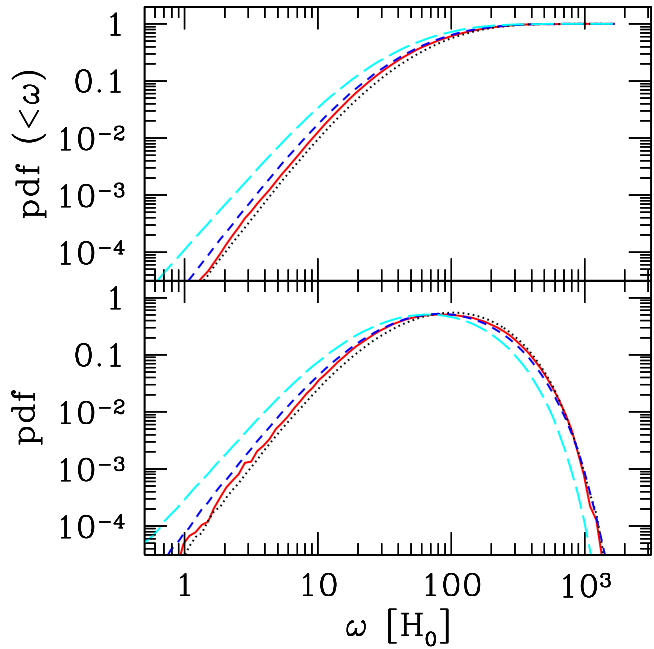}
\caption{Cumulative (top) and ordinary (bottom) probability density
function of the vorticity per Log vorticity interval.
Vorticity is computed on a scale
twice the mesh size of the finest grid, i.e. 
$\ell=2\Delta x_5=14.6\,h^{-1}$ kpc.
Different lines from left to right correspond to 
the following regions: off-virial for
$R_\mathrm{vir}<R<\frac{4}{3}R_\mathrm{vir}$ (long-dash cyan),
virial radius for $\frac{2}{3}R_\mathrm{vir}<R<R_\mathrm{vir}$ (short-dash blue),
off-core for $\frac{1}{3}R_\mathrm{vir}<R<\frac{2}{3}R_\mathrm{vir}$ (dot black),
core for $R<\frac{1}{3}R_\mathrm{vir}$ (solid red).
Note how dot and solid lines are nearly indistinguishable.
\label{t_f9:fig}}
\end{figure}

The panel on the left hand side of Figure~\ref{t_f7:fig} shows the
distribution of the vorticity in a plane passing through the GC
center. The plane orientation is as in the middle panels of
Figure~\ref{t_f7:fig}, although the distribution of vorticity on different
planes is quite similar.  The panel on the right hand side is a zoom
in of the shaded area in the central region of the left panel.  The
vorticity is evaluated on scales scales $\ell=2\Delta
x_5=14.6\,h^{-1}$ kpc, corresponding to the finest grid, with a
cell-centered finite difference scheme, and is expressed in units of
$H_0$.

Figure~\ref{t_f7:fig} shows that inside the accretion shocks the value
of the vorticity is quite high, ranging from a few tens (yellow) to a
few thousand (red). This is consistent with results in, e.g.,~\cite{Ryu08}
and~\cite{Zhu11}.  Since vorticity is expressed in units of $H_0$,
this means that vortical motions are fully developed on scales of a
few tens of kpc.  The right panel of Figure~\ref{t_f7:fig} shows more
clearly that the vorticity developed very fine structure, all the way
down to the finest scales allowed by the mesh size of the grid.

Figure~\ref{t_f8:fig} presents a baroclinic term versus vorticity
phase-space diagram.  The three panels from left to right correspond
to the core, off-core + virial, and off-virial regions, respectively.
The peak value of both vorticity and baroclinic term become stronger
in the outer regions of the cluster, in agreement with
Figure~\ref{t_f5:fig} and ~\ref{t_f7:fig}.  In addition, this figure
also shows a weak correlation between the two quantities.  Namely the
baroclinic term appears roughly proportional to the square of the
vorticity, a relation followed by oblique dash lines (with the thick
(red) line corresponding to unit proportionality constant). Most of
the vorticity lives in a region of phase space where the constant of
proportionality is between 1 and 10$^{-2}$.  This is suggestive that
the vorticity develops on a time scale roughly between 1-100 times its
inverse.  A similar qualitative conclusion can be inferred by the
analysis of the dashed lines in Figure~\ref{t_f3:fig}, which
correspond to curves $\dot\omega_{\rm baroclinic}\propto (\nabla\cdot
v)^2$, if we replace the vorticity with the velocity divergence.

The top panel of Figure~\ref{t_f9:fig} shows the cumulative
probability density function (pdf) of the vorticity per $\log$
vorticity interval.  Different curves correspond to different
spherical shells around the GC, i.e. core (solid-red line), off-core
(dotted-black line), virial (short-dashed-blue line), and finally,
off-virial (long-dashed-cyan line). The three pdf's within the virial
radius (solid, dot and short-dash lines) are almost indistinguishable
while the peak of the long-dashed line (off-virial) is shifted to the
left by a factor $\lesssim 2$ compared to solid line (core),
suggesting a slight trend for the vorticity to become slower, on
average, towards
the outer regions.  In comparison, inspection of Figure~\ref{t_f9:fig}
indicates that the vorticity outside the accretion shocks is much
lower and is narrowly distributed around values a few times 0.1 $H_0$.
In any case, the pdf's in Figure~\ref{t_f9:fig} confirm the visual
impression provided by the two-dimensional slice that high vorticity
is characterized by a quite large volume filling factor. For example,
99\% of space inside the virial volume has vorticity $\omega> 10
H_0$. Outside the virial radius, that percentage drops only slightly
to 96\%.  The high filling factor of the vorticity found here is in
contrast with the results in~\cite{Iapichino08}, most likely because
the GC studied by those authors was a relatively relaxed system.

Since the numerical viscosity acts on scales larger than the actual
collisional mean free path, the Reynolds number of the simulated flow
is not sufficient to ensure high filling factor of the turbulent wakes
generated by moving galactic substructure~\citep{Subramanian06}.
Therefore, this suggests that most of the vorticity inside the ICM is
generated by shocks and filaments.

If we assume Kolmogorov scaling for the vorticity, $\omega_\ell\propto
v_\ell/\ell\propto\ell^{-2/3}$, we can estimate the pdf of the
vorticity also on larger scales by shifting accordingly the x-axis of
Figure~\ref{t_f9:fig}, i.e. by a factor $(10^3/14.6)^{2/3}\simeq 26$ for
scales comparable to the injection scale.  We then conclude that even
on the largest scales where we expect shear flows to be generated, the
vortical motions remain fast compared to the Hubble time, for a
significant fraction of the GC volume and at different distances from
the GC center.

\begin{figure}[t]
\centering
\includegraphics[width=0.5\textwidth]{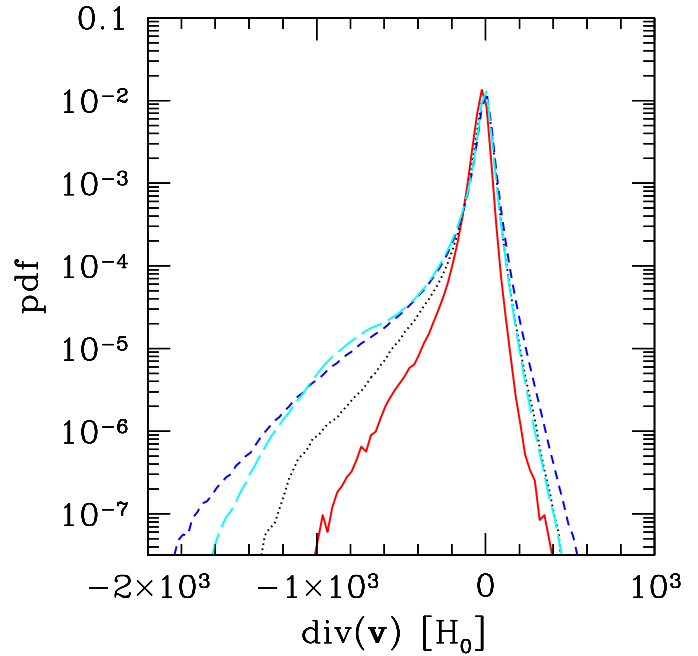}
\caption{Probability distribution function of the velocity 
divergence. Line style and color is the same as in Figure~\ref{t_f9:fig}.
\label{t_f10:fig}}
\end{figure}
\begin{table}  [t]
\begin{center}
\begin{small}
\caption{pdf's Fit Parameters \label{omepdf:tab}}
\begin{tabular}{cccccc}
\hline
\hline
{Region} &
{$\bar\omega$} &
{$\sigma_\omega$}
 \cr
 & ($H_0$) &  ($H_0$) \cr
\hline
core      & 14 & 240  \cr
off-core  & 16.5 & 240  \cr
virial    & 12 & 240   \cr
off-virial& 9 & 185  \cr
\hline
\hline 
\end{tabular}
\end{small}
\end{center}
\end{table}
The bottom panel of Figure~\ref{t_f9:fig} shows the (non-cumulative)
vorticity pdf per $\log$ interval.  We find that each pdf can be very
well approximated by function of the form
\begin{equation}\label{fit:eq}
\mathrm{pdf} = A 
\frac{\omega^3}{\omega^2+\bar\omega^2}
e^{-\frac{|\omega-\bar\omega|}{\sigma_\omega}},
\end{equation}
where, $\bar\omega$ and $\sigma_\omega$ are fitting parameters but can
be identified as is the peak value and the width of the distribution,
respectively, while $A$ is a normalization factor.  Fitting parameters
for the various curves are reported in Table~\ref{omepdf:tab}.
In particular, for large values of the vorticity, $\omega>\bar\omega$,
the pdf's exhibit an exponential cutoff. This feature is characteristic
of fully developed compressible	turbulence with modest Mach number,
as demonstrated in the numerical simulations in~\cite{Porter02}.

The same features characteristic of fully developed compressible
turbulence appear in the pdf of the velocity divergence presented in
Figure~\ref{t_f10:fig}.  Different curves are style and color coded as
in Figure~\ref{t_f9:fig}.  In particular, the strong asymmetry of the
pdf and the extended wings towards negative value of the divergence
with almost power-law shape, is associated to the presence of numerous
shocks and resemble qualitatively those reported in~\cite{Porter02}
and~\cite{Schmidt09}.
\begin{figure}[t]
\centering
\includegraphics[width=0.5\textwidth]{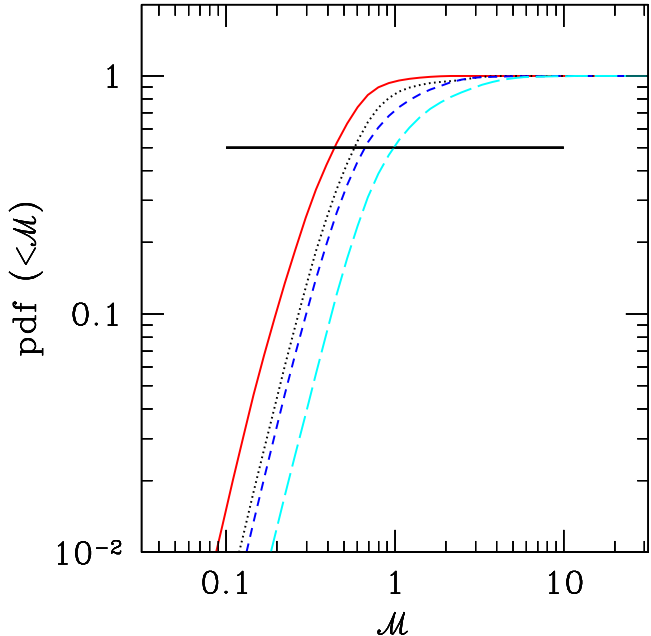}
\caption{Cumulative probability distribution function of the flow 
Mach number for different regions around the GC center.
Line style and color is the same as in Figure~\ref{t_f9:fig}. The horizontal
line corresponds to the value $\frac{1}{2}$, intersecting each pdf
at the median value.
\label{t_f11:fig}}
\end{figure}

\begin{deluxetable*}{lccccccccccccc}
\tabletypesize{\small}
\tablecaption{Turbulence Results\label{turb:tab}}
\tablecolumns{14}
\tablewidth{0pt}
\tablehead{
\colhead{Region} & 
\multicolumn{6}{c}{solenoidal}    & &
\multicolumn{6}{c}{compressional}
\\ \\
\cline{2-7} \cline{9-14} 
\\ \\
\colhead{} &
\colhead{$L^\parallel_{\rm inj}$} &
\colhead{$\zeta^\parallel_2$} &
\colhead{$\zeta^\parallel_3$} &
\colhead{$L^\perp_{\rm inj}$} &
\colhead{$\zeta^\perp_2$} &
\colhead{$\zeta^\perp_3$} &
\colhead{} &
\colhead{$L^\parallel_{\rm inj}$} &
\colhead{$\zeta^\parallel_2$} &
\colhead{$\zeta^\parallel_3$} &
\colhead{$L^\perp_{\rm inj}$} &
\colhead{$\zeta^\perp_2$} &
\colhead{$\zeta^\perp_3$} \\
\colhead{} &
\colhead{($R_{\rm vir}$)} & \colhead{} & \colhead{} &
\colhead{($R_{\rm vir}$)} & \colhead{} & \colhead{} &
\colhead{} &
\colhead{($R_{\rm vir}$)} & \colhead{} & \colhead{} &
\colhead{($R_{\rm vir}$)} & \colhead{} & \colhead{} 
}
\startdata
core       & $\approx 0.7$& 0.80& 1.15& $\approx 0.6$& 0.69& 0.97& & $\approx 0.9$& 1.16& 1.56& $\approx 0.8$& 1.24&1.61 \cr
off-core   & $\approx 1.0$& 0.67& 0.95& $\approx 0.8$& 0.64& 0.91& & $>1.0$& 0.97& 1.20&$\approx 0.8$& 1.10&1.30 \cr
virial     & $\approx 1.5$& 0.60& 0.82& $\approx 1.0$& 0.58& 0.78& & $>1.0$& 0.77& 0.82&$\approx 0.9$&1.06&1.17 \cr
off-virial &$>1.5$& 0.66& 0.91& $\approx 1.2$& 0.61& 0.83& &$>1.5$& 0.85& 0.95&$>1.5$& 0.98&1.05 
\enddata
\end{deluxetable*}

Finally, in Figure~\ref{t_f11:fig} we plot the cumulative pdf of the
Mach number, $\mathcal{M} \equiv u/c_s$, with $c_s$ the gas sound
speed.  As in the previous figures, the four curves in
Figure~\ref{t_f11:fig} correspond to the four regions around the GC
center, from the core (leftmost, solid red curve) to the off-virial
region (rightmost, long-dash cyan curve).  Since the gas is not
isothermal, with a general trend of the temperature to decrease
towards the outer regions (see Figure~\ref{t_f2:fig}), the Mach number
is a function of both the fluid element velocity and local
temperature.  The intersection of the pdf with the solid horizontal
line marks the median Mach number of the pdf. In fact, the cumulative
pdf's in Figure~\ref{t_f11:fig} become broader towards the outer
region, indicating that the flow is progressively more compressible.
The higher level of turbulent energy found in the low density
intergalactic medium compared to the ICM by~\citet{Iapichino11} is
probably related to this effect. In the core region, the Mach number is
mostly between 0.1-1, with a median value of $\lesssim 0.5$,
indicating that the gas is mildly compressible. Moving outward in
space the gas compressibility increases, with the median Mach number
taking the values $0.6$ (off-core), $0.7$ (virial) and $\sim 1$
off-virial.

In conclusion, the analysis carried out in this section indicates that
both in the core and in the distant outskirts the vorticity is volume
filling and with a timescale short compare to the Hubble time even on
scales comparable to $R_\mathrm{vir}$. In addition, the pdf of the
vorticity and the velocity divergence are qualitatively very similar
to those observed in simulations of compressible turbulence with
comparable Mach numbers to those characterizing our simulated flows.
The median Mach number is typically less than one, ranging from 0.5
to 1 from the core to the off-virial regions, respectively.  Consistent with
expectations based on the analysis in Sec.~\ref{tim:sec}, the ICM
appears in a state of fully developed turbulence.  
This conclusion is further corroborated by results in the next section.

\subsection{Structure Functions}\label{sf:sec}
\begin{figure*}[t]
\centering
\includegraphics[width=0.5\textwidth]{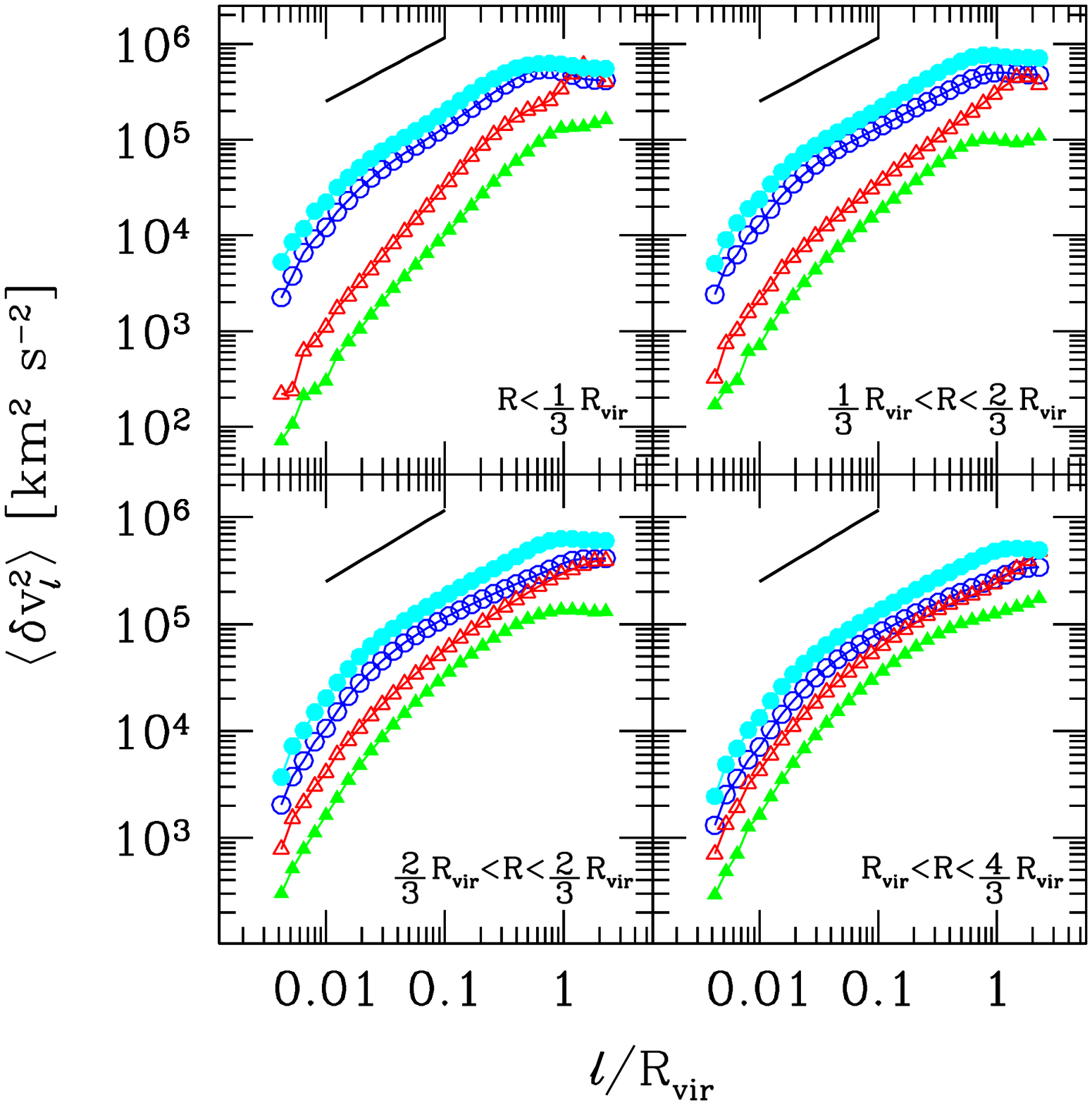}\includegraphics[width=0.5\textwidth]{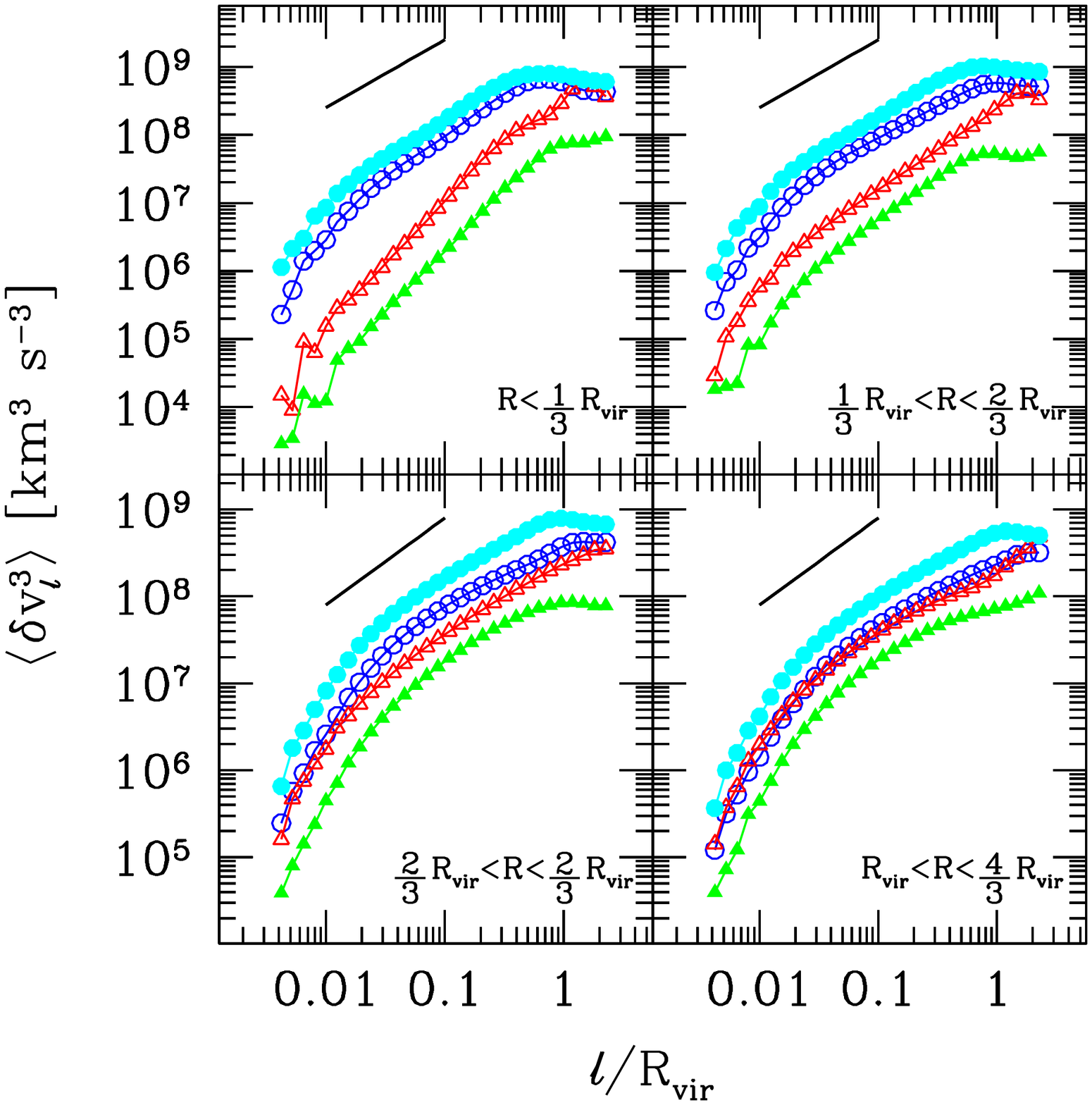}
\caption{Second (left) and third (right) order structure function of
  the velocity field.  Different curves from top to bottom correspond
  to the following components: solenoidal--longitudinal (blue open
  circles), solenoidal--transverse (cyan solid circles),
  compressible--longitudinal (red open triangles),
  compressible--transverse (green solid triangles).  The structure
  functions in the different panels were sampled from different shell
  volumes centered around the GC center, namely: core corresponding to
  $R<(1/3)R_\mathrm{vir}$ (top left), off-core corresponding to
  $(1/3)R_\mathrm{vir}< R<(2/3)R_\mathrm{vir}$ (top right), virial
  corresponding to $(2/3)R_\mathrm{vir}< R<R_\mathrm{vir}$ (bottom
  left), off-virial corresponding to
  $R_\mathrm{vir}<R<(4/3)R_\mathrm{vir}$ (bottom right).
\label{t_f12:fig}}
\end{figure*}
In this section we study the statistical properties of the velocity
increments as a function of spatial separation. This is carried out
by computing the structure functions of order $p$, for both the
longitudinal and transverse components of the velocity field.  In
addition, the velocity field is decomposed into the solenoidal,
$\vvec_s$, and potential or compressional, $\vvec_c$, components using the
Hodge-Helmholtz decomposition, i.e.
\begin{gather} \label{hh0:eq}
\vvec=\vvec_s+\vvec_c, \\
\vvec_c=-\nabla\phi, \quad \vvec_s=\nabla\times \mathbf{A}, \\
\phi =\frac{1}{4\pi}\int \frac{\nabla\cdot\vvec}{r} d\xv, \quad
\mathbf{A} =\frac{1}{4\pi}\int \frac{\nabla\times\vvec}{r} d\xv. \label{hh2:eq}
\end{gather}
The longitudinal and transverse structure functions of order $p$ are
then computed for each velocity component as
\begin{eqnarray} \label{lsf:eq}
\langle \delta \vvec_\ell^p \rangle _{l,q} &=& 
\langle |[\vvec_q(\xv+\ell\nv)-\vvec_q(\xv)]\cdot \nv|^p \rangle, \\ \label{tsf:eq}
\langle \delta \vvec_\ell^p \rangle _{t,q} &=& 
\langle |[\vvec_q(\xv+\ell\nv)-\vvec_q(\xv)]\cdot \tv|^p \rangle,
\end{eqnarray}
where $\ell$ is the spatial separation and $\nv,~\tv$ are a unit
directional vector, with $\tv\equiv(\nv\times
(\vvec_q\times\nv))/|\vvec_q|$, while $q=c,s$, refers to the solenoidal and
compressional components, respectively.

In Figure~\ref{t_f12:fig} we present second (left) and third order
(right) velocity structure functions for the different velocity
components discussed above and extracted from different sub-volumes of
the GC.  In particular, curves from top to bottom correspond to
solenoidal-transverse (cyan circles), solenoidal-longitudinal (blue
circles), compressible-transverse (red triangles) and
compressible-longitudinal (green triangles) components, respectively.
In addition, the different panels correspond to the core (top left),
off-core (top right), virial (bottom left), and off-virial (bottom
right) region, respectively.  Finally, for reference, the black line
shows the prediction for fully developed incompressible, isotropic and
homogeneous turbulence ~\citep[][hereafter
K41]{Kolmogorov41a,Kolmogorov41c,Oboukhov41a}.
To compute the structure function we define sampling points randomly
distributed inside the volume of interest and compute the velocity
difference with respect to randomly selected field points at a maximum
distances of 2 $R_\mathrm{vir}$. A total of $\lesssim 10^6$ sampling
points and $\lesssim 10^6$ field points are used for the purpose,
assigned to each sub-region proportionally to volume.

A number of important features characterize the structure functions in
Figure~\ref{t_f12:fig}.  On scales $\ell\lesssim \alpha_l$\rvir, with
$\alpha_l\sim 3-4\times 10^{-2}$, corresponding to $\ell \simeq
10\times\Delta x\simeq 100$ kpc, the structure functions generally
steepen, due to the influence of numerical dissipation as discussed in
Sec.~\ref{app:sec}.  At separation scales $\ell\gtrsim\alpha_h$\rvir,
with $\alpha_h\sim 1$ the velocity increments are no more correlated
and the structure functions tend to flatten out, although non-zero
slope is noticeable most likely as a result of spatial
inhomogeneity. The scale $\alpha_h$\rvir~grossly defines the curvature
radius of the accretion and internal shocks, as well as the large
scale shear flows, as discussed in Sec.~\ref{tim:sec}.  Furthermore,
as discussed in Sec.~\ref{fdt:sec}, the timescale associated to
vortical motions at this scale are short compared to the Hubble time.
Therefore this is most likely the injection scale of the turbulence,
whereas the scale interval
$\alpha_l\lesssim\ell/R_\mathrm{vir}\lesssim\alpha_h$, where both the
second and third order structure functions exhibit power-law scale
free behavior, $\langle \delta \vvec_\ell^p \rangle\propto
\ell^{\zeta_p}$, is identified as the inertial range of the turbulent
cascade.  For comparison, a similar value for the injection scale was
found by~\citet{Vazza11} who computed the third order structure
function, averaged over a sample of clusters, for the total velocity
($\vvec_s+\vvec_c$) and the whole GC volume, without longitudinal or
transverse projection~\citep[see also][for density weighted structure functions in the core region]{Valdarnini11}.

The properties of the structure functions reported in
Figure~\ref{t_f12:fig}, including the injection scales and the slopes
in the inertial range, are summarized in Table~\ref{turb:tab}.  These
numbers should not be taken as precise ``measurements'', because
occasionally the structure functions themselves are not perfect power
laws even in the inertial range. In addition, the condition of the ICM
are changeable, so the values determined at this particular time and
for this particular system will be different for different times and
systems. However, the qualitative properties reflected in the reported
value have probably general character and are discussed below.
Similarly, the value of the velocity increments at given separation
scales can be compared to the sound speed as inferred from the
temperature profile provided in Figure~\ref{t_f2:fig}, to check for
departure from conditions of hydrostatic equilibrium. The same
cautionary note, however, applies here.

For both the solenoidal and compressional components the injection
scale of the turbulence appears to increase towards the GC outer
region, ranging from $\gtrsim 0.7$\rvir~in the core to the
$\gtrsim$\rvir~around the virial region. Note that, particularly in
the outer GC regions, the longitudinal structure functions do not
flatten as much as the transverse component.  This is likely due to
the fact that as we approach the virial region, we start to sample
velocity differences with respect to gas outside the external
accretion shocks, where the flows is laminar.
As for the solenoidal component, the spectral slope in the inertial
range is close to the value predicted by Kolmogorov's theory for
incompressible, homogeneous and isotropic
turbulence~\citep{Kolmogorov41a,Kolmogorov41c,Oboukhov41a}, despite
the strong departures from the assumptions in that theory. This is in
agreement with previous work on compressible turbulence with moderate
to large Mach numbers~\citep{Porter92,Porter02,Kritsuk07}.  Note that
the spectral slopes become slightly flatter towards the outskirt
region of the GC.  This is related to the corresponding increase in
the turbulence driving scale mentioned above and is discussed further
below.  The spectral slopes of the structure functions for the
compressible components (open and solid triangles), are significantly
steeper than in the solenoidal case, particularly in the core region,
and actually roughly consistent with Burgers' model of
turbulence~\citep{Burgers39}. Note, however, that as for the
solenoidal case the compressible structure functions also tend to
become flatter towards the outer region of the GC.

\begin{figure}[t]
\centering
\includegraphics[width=0.5\textwidth]{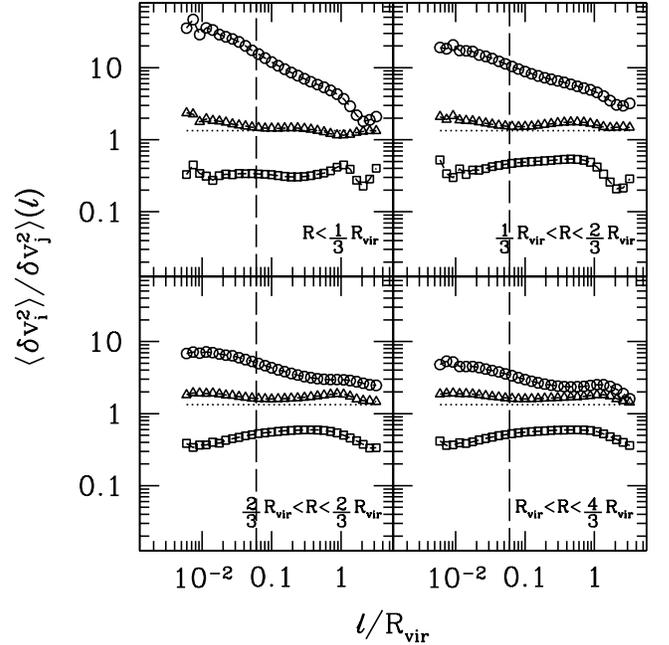}
\caption{Ratio of second order velocity structure functions:
solenoidal-transverse to solenoidal-longitudinal (triangle),
compressional-transverse to compressional-longitudinal (squares),
and total solenoidal to compressional components (circles), respectively.
The dot horizontal line indicates a ratio 0.5, and the vertical dash line
a length of 15 resolution elements.
The four different panels correspond to the same region around the GC
center as in Figure~\ref{t_f12:fig}.
\label{t_f13:fig}}
\end{figure}
For incompressible, isotropic and homogeneous turbulent
flow, an analytic relation exists between the transverse and
longitudinal second order structure
functions~\citep{vonKarmanHowarth38,LandauLifshitz6}, i.e.
\begin{equation}\label{vKH:eq}
\langle \delta \vvec_\ell^2 \rangle _{t,s} =\frac{2+\zeta_2}{2}\langle \delta \vvec_\ell^2 \rangle _{l,s}\approx \frac{4}{3}\langle\delta\vvec_\ell^2 \rangle _{l,s}
\end{equation}
where the last approximation assumes close to K41 scaling.  This
relation holds to a good approximation even for compressional flows,
as demonstrated in~\cite{Kritsuk07}, who carried out simulations of
isothermal supersonic turbulence with rms Mach number $\simeq 6$.  In
accord with the prediction from Eq. (\ref{vKH:eq}) the transverse and
longitudinal structure functions share a very similar scaling exponent
and their ratio is close to 4/3. This is illustrated in more detail in
the four panels of Figure~\ref{t_f13:fig}, where the ratio is plotted
explicitly (triangles) as a function of spatial separation and the
prediction by~\cite{vonKarmanHowarth38} is illustrated by the
horizontal dotted line.  In this figure the vertical line correspond
to a separation of 16 resolution elements, roughly the scale where
numerical dissipation becomes appreciable. The figure shows that in
the core region, on scales where the turbulence motions are resolved,
the solenoidal flow is consistent with homogeneous and isotropic
turbulence. Anisotropy gradually arises towards the outer regions, and
particularly close to the injection scales.  This could be related to
the fact that in the outer regions the flow becomes increasingly
compressional, i.e. the growing presence of shocks which inject
turbulence on various scales through the baroclinic term, on
timescales comparable to those required to reach isotropization.  Note
also that anisotropy appears on scales where the turbulence cascade is
not resolved, most likely because the longitudinal compression suffers
excessive numerical dissipation, which does not affect the relative
transverse motions. This means care must be taken when studying
anisotropy of turbulent motions on small scales, because it is in
general affected by numerical artifacts.

In Figure~\ref{t_f13:fig} we also show the ratio of transverse to
longitudinal second order structure functions for the compressible
velocity (open squares). To the best of our knowledge there is no
analytic prediction for this quantity.  This ratio is much smaller
than for the solenoidal case and it appears to be close to $\half$,
particularly in the outer regions.  In other words, the longitudinal
term largely dominates the compressional velocity components, which is
not surprising because unlike the solenoidal case, longitudinal
compression now leads to dissipation as opposed to excitation of
transverse modes.

\begin{figure*}[t]
\centering
\includegraphics[width=0.5\textwidth]{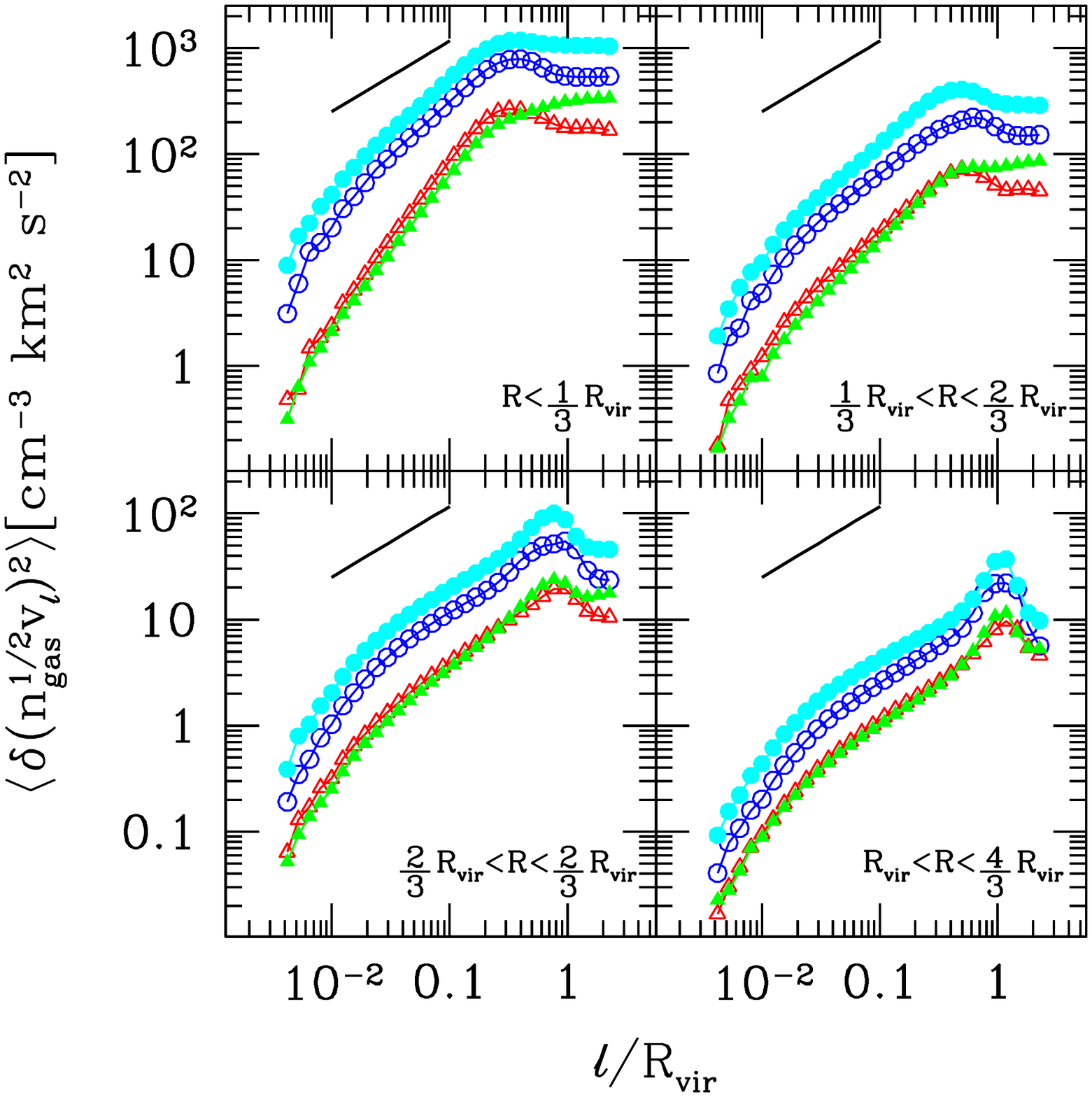}\includegraphics[width=0.5\textwidth]{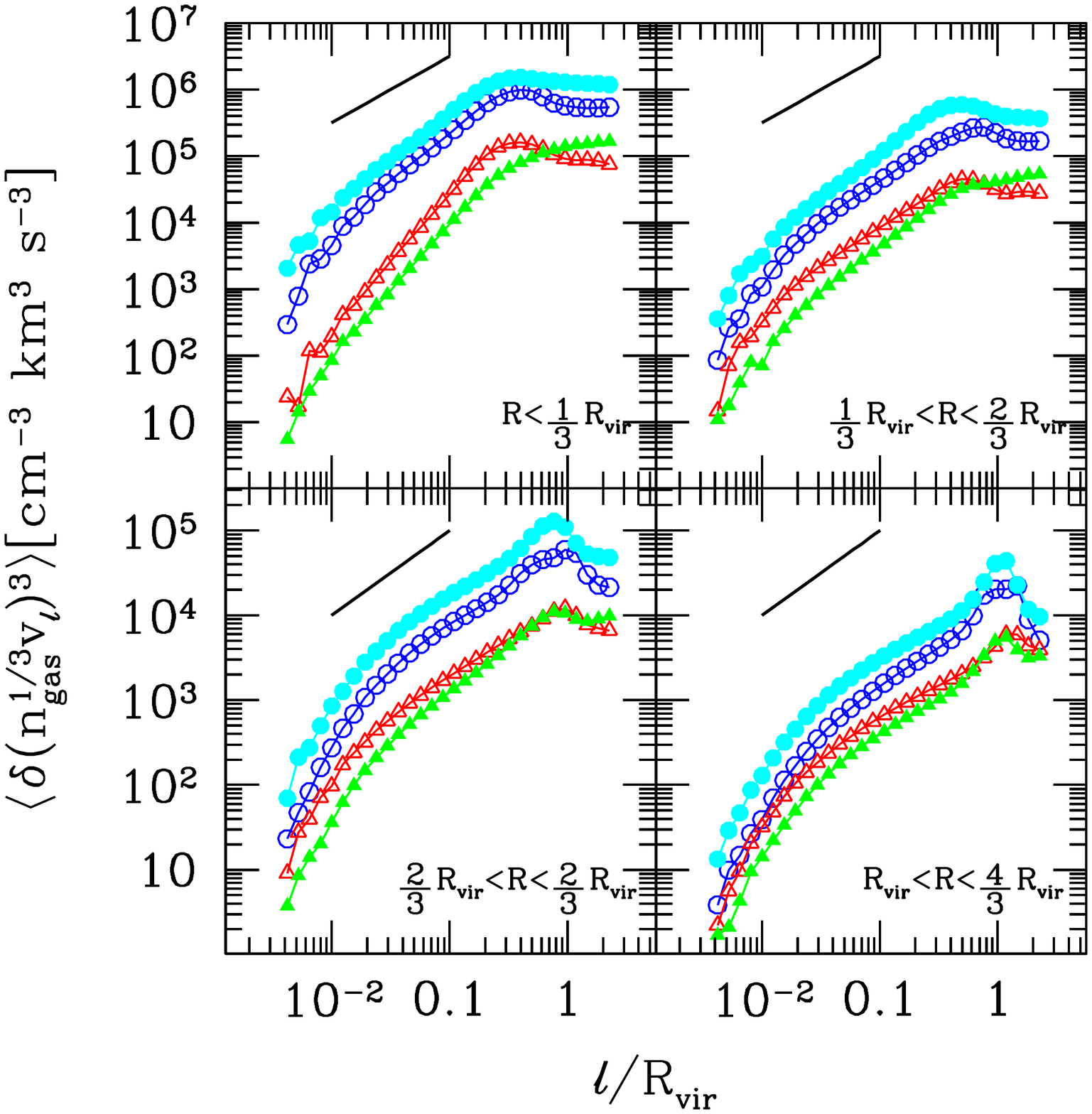}
\caption{
Density weighted second (left) and third (right) order structure function of the velocity field. 
Different curves from top to bottom correspond to the following components: 
solenoidal-longitudinal (blue circles), solenoidal-transverse (cyan circles), 
compressible-longitudinal (red triangles), compressible-transverse (green triangles). 
The structure functions in the different panels were sampled from
different shell volumes centered around the GC center, namely:
core corresponding to $R<(1/3)R_\mathrm{vir}$ (top left), off-core
corresponding to $(1/3)R_\mathrm{vir}< R<(2/3)R_\mathrm{vir}$ (top right),
virial corresponding to $(2/3)R_\mathrm{vir}< R<R_\mathrm{vir}$ (bottom left),
off-virial corresponding to $R_\mathrm{vir}<R<(4/3)R_\mathrm{vir}$ (bottom right).
\label{t_f14:fig}}
\end{figure*}
Finally, note that the turbulent energy is generally dominated by the
solenoidal component.  This is evident from the plot in
Figure~\ref{t_f12:fig}, and is again illustrated explicitly as a function
of separation scale by the open circles in Figure~\ref{t_f13:fig} for
different regions across the GC.  In the core region,
$R\leq\frac{1}{3}$\rvir, and at separation scales
$\ell\sim\frac{1}{2}$\rvir, the compressional component accounts for
about 5\% of the total turbulent energy.  This value is of course
scale dependent, and it rises to a value of order a few close to the
injection scale, but drops further towards smaller scales, as a result of
the steeper scaling of the compressional versus solenoidal components
of the turbulent velocity.  However, in the outer region the above
ratio at fixed separation scale generally increases, so again at
$\ell\sim\frac{1}{2}$\rvir, it reaches values about 20\% at the virial
and off-virial regions.

We have also carried out the same analysis for density weighted second
and third order velocity structure functions, where the velocity is
again decomposed according to solenoidal and compressional components.
Basically we use the same equations~(\ref{hh0:eq})-(\ref{hh2:eq}), but
with the following substitution $\vvec_q \leftarrow \rho^\alpha
\vvec_q$, with $\alpha=1/p$ and $p$ the order of the structure
function.  This choice of the density weighting corresponds to recent
attempts to generalize Kolmogorov's scaling for incompressible to
compressible case~\cite[e.g.][]{Kritsuk09}.  In any case, the results
for the structure function of second and third order are shown in
Figure~\ref{t_f14:fig}, where the black solid line and the different
curves and panels have the same meaning as in Figure~\ref{t_f12:fig}.
Basically the density weighted structure functions are consistent with
the results presented in Figure~\ref{t_f12:fig} for the simple
structure functions.  If anything, the outer scale where the structure
functions flatten and turn over is smaller by a factor $\lesssim 2$
and the slope in the inertial range, although still consistent with
K41, is slightly steeper than in the corresponding cases illustrated
in Figure~\ref{t_f12:fig}.  However, the structure functions in this
case show mild structure both in the inertial range and at the outer
scales. This effect, which is most likely attributable to density
inhomogeneity, is negligible within $\frac{2}{3}$\rvir, but becomes
evident for the virial and off-virial regions.
\subsubsection{Convergence}\label{conv:sec}
\begin{figure*}[t]
\centering
\includegraphics[width=0.5\textwidth]{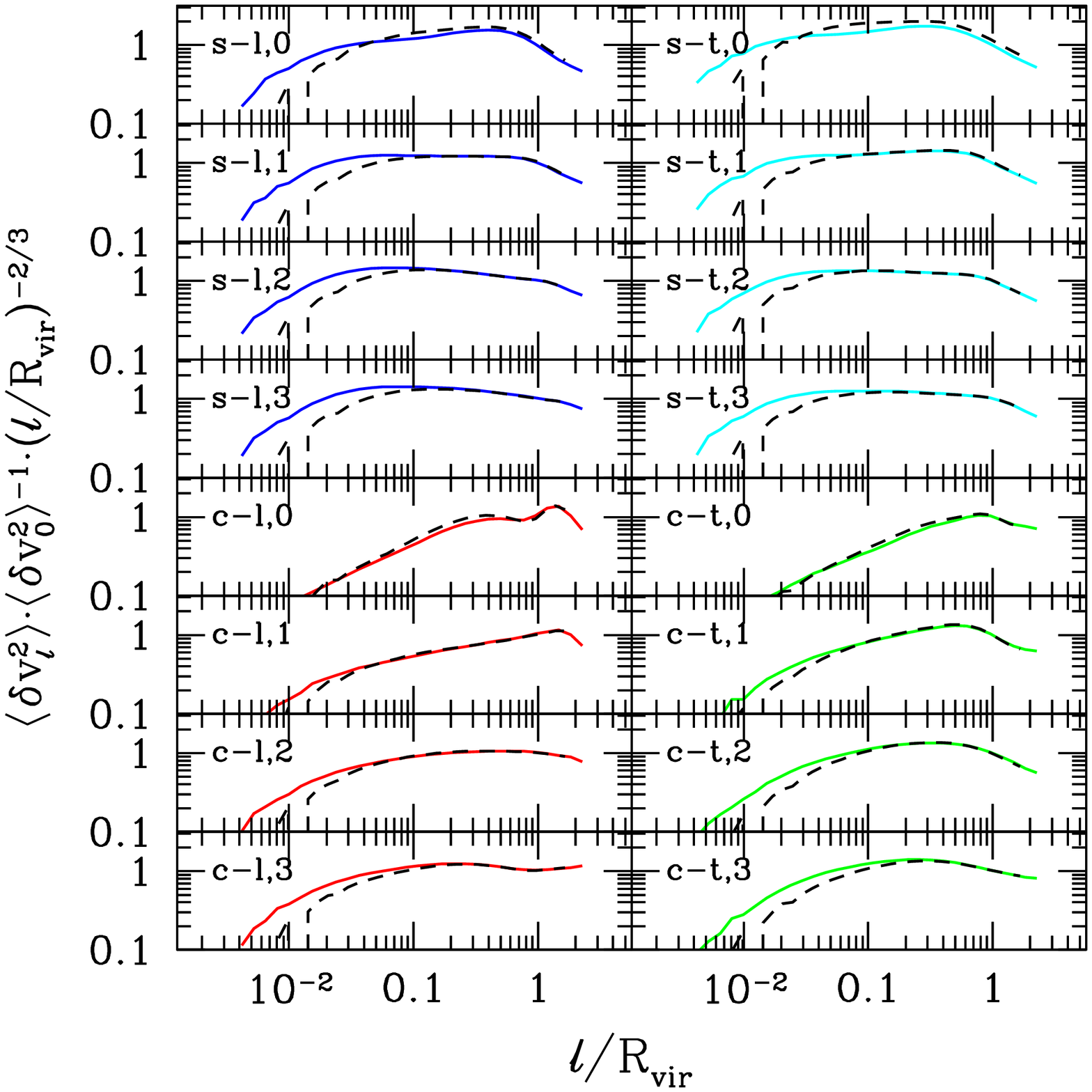}\includegraphics[width=0.5\textwidth]{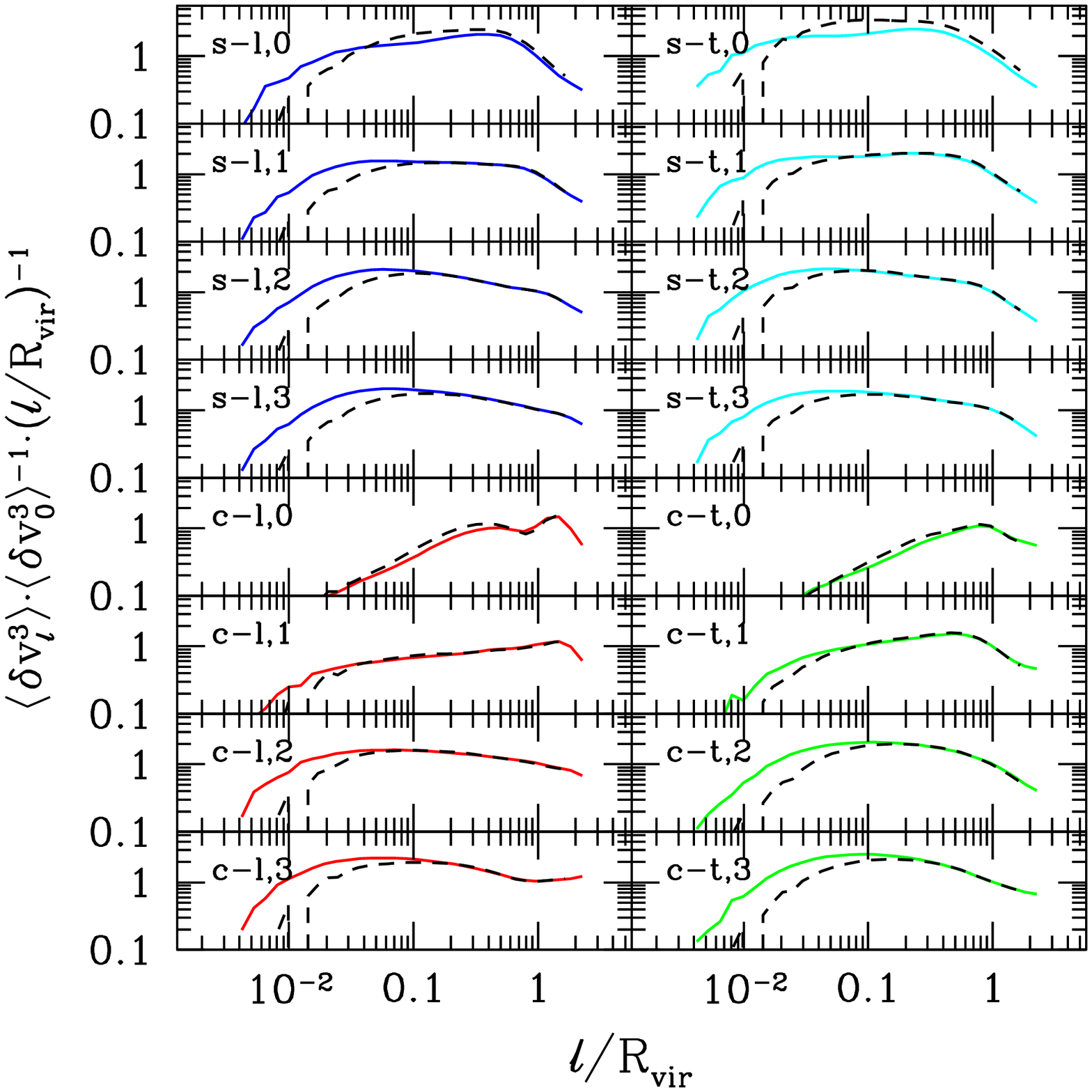}
\caption{
 Compensated second (left) and third (right) order structure function of the velocity field from our fiducial run (solid line) and a run at lower resolution by 
a factor 2 (dashed line). The quantity $\delta v_0$ entering the normalization
factor corresponds to the velocity increment on scales $\simeq$\rvir, as measured in the fiducial run.
Labels s-l,s-t,c-l,c-t indicate, respectively, solenoidal-longitudinal (blue), solenoidal-transverse (cyan), compressible-longitudinal (red), and compressible-transverse (green) component.
Labels 0,1,2,3 indicate, respectively,
core ($R<(1/3)R_\mathrm{vir}$), off-core
($(1/3)R_\mathrm{vir}< R<(2/3)R_\mathrm{vir}$),
virial ($(2/3)R_\mathrm{vir}< R<R_\mathrm{vir}$),
and off-virial ($R_\mathrm{vir}<R<(4/3)R_\mathrm{vir}$) region.
\label{t_f15:fig}}
\end{figure*}

The convergence of numerical results related to the statistical
properties of turbulence is well documented in studies of steady-state
fully developed turbulence in periodic
boxes~\citep{Porter92,Porter94,Federrath11}.  Once steady-state
conditions are achieved, the inertial range of the turbulent cascade
emerges, provided there is enough separation between the injection and
the dissipation scale (e.g., a factor ten).  While the former is set
by the driving mechanism, and typically acts on scales comparable to
the computational domain size, the latter is determined by the
dissipation properties of the numerical scheme.  In the case of the
PPM method, which is commonly employed for compressible turbulence,
numerical dissipation affect the turbulent cascade up to separation
scales of 32 resolution elements. As a result, on scales of this order
and smaller, the turbulence cascade steepens.  On larger scales, the
cascade is determined by the energy transfer due to nonlinear fluid
interactions. As the resolution is increased, numerical dissipation
occurs on smaller spatial scales.  Therefore, the inertial range of
the turbulence cascade where the nonlinear effects dominate extends to
smaller separation scales, while on larger scales its statistical
properties remain unchanged (structure function or power spectra). In
reality relatively small differences may appear due to temporal
fluctuations, particularly on large scales with relatively lower
number of degrees of freedom. For this reason, time averaging is also
employed.

Guided by the above studies, we try and assess
the convergence of the statistical properties of the turbulence in our
calculation, by carrying out a lower resolution run which employs 4
instead of 5 levels of refinement. So in this case the resolution
inside the GC volume is a factor of two lower.  The results are
summarized in Fig.~\ref{t_f15:fig}.  Here we compare the structure
functions from the low resolution run (dashed line) with those
obtained from the fiducial run (solid).  We compare by component and
by region. So labels s-l,s-t,c-l,c-t in the top-left corner of each subpanel
correspond, respectively, to the components: solenoidal-longitudinal
(blue), solenoidal-transverse (cyan), compressible-longitudinal (red),
compressible-transverse (green); while the labels 0,1,2,3, indicate,
respectively, the region: core, off-core, virial and off-virial.
Second and third order structure functions are presented in the left
and right panel, respectively. For the sake of clarity, structure
functions are (partially) compensated, i.e., multiplied by a factor
$(\ell/R_{\rm vir})^{p/3}$, where $p$ is the order, and normalized by
a factor, $\langle\delta v_0^p\rangle$, where $\delta v_0$ is the velocity
increment on scales $\simeq$\rvir, as measured in the fiducial run.

In the low resolution run, 32 resolution elements correspond to a
scale, $\ell/$\rvir$\simeq 0.24$, in Figure~\ref{t_f15:fig}.
Inspection of the various panels in Figure~\ref{t_f15:fig} show that
in almost all the cases the structure functions from the low
resolution and the fiducial run are in good agreement at least on
scales a factor two smaller that the one quoted above.  This suggests
that the fiducial run has definitely reached convergence at the
nominal scale of 32 resolution elements, $\ell/$\rvir$\simeq 0.12$,
but that the results are probably fairly reliable also on smaller
scales down to $\ell/$\rvir$\simeq 0.06$.

Perhaps the one exception is for the structure functions of the
solenoidal-longitudinal components (s-l) in the core region (0). Here
one can observe extra power in the low resolution run compared to the
fiducial run, particularly in the third order statistics. This is
possibly due to the well known bottleneck effect, whereby the energy
transfer due to nonlinear effects becomes inefficient below a certain
scale, causing accumulation of kinetic energy on larger scales
thereof.  To fully address the issue, one would require a calculation
with higher resolution than in our fiducial case. We postpone this
task to future work. Nevertheless, in terms of convergence of the
fiducial run, the above bottleneck effect appears to affects scales
smaller than $\ell/$\rvir$\simeq 0.1$.  Intriguingly, the same effect
does not appear in the same or any other structure functions computed
in the outer parts of the GC volume.  If the bottleneck effect is
indeed the culprit, one can surmise that its impact becomes less
important in the outer regions of the GC, because there more efficient
dissipation is provided by the increasing compressional component of
the turbulence, which decays through weak shocks.

\subsubsection{Comparison with Lagrangian AMR}\label{amrcomp:sec}
\begin{figure*}[t]
\centering
\includegraphics[width=0.5\textwidth]{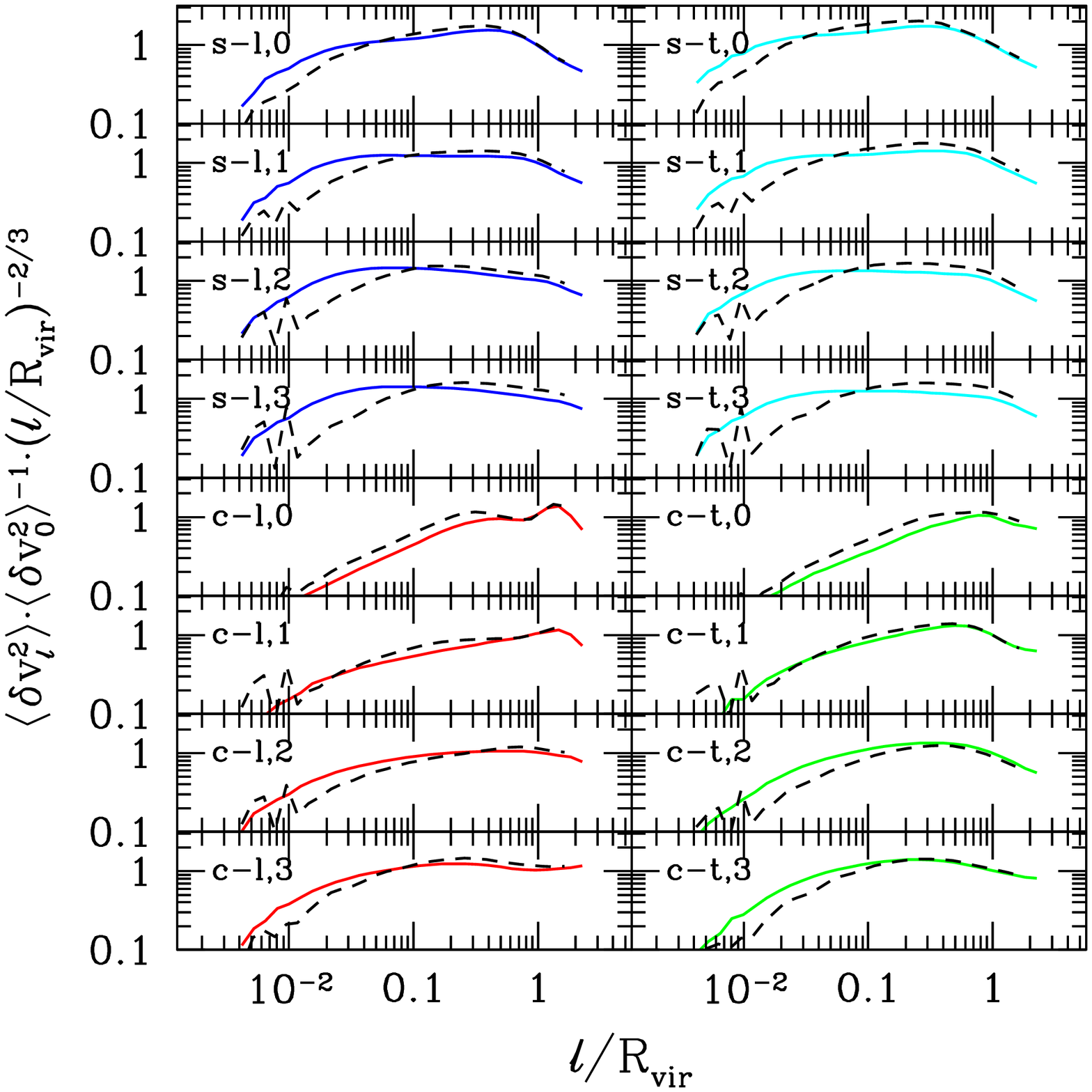}\includegraphics[width=0.5\textwidth]{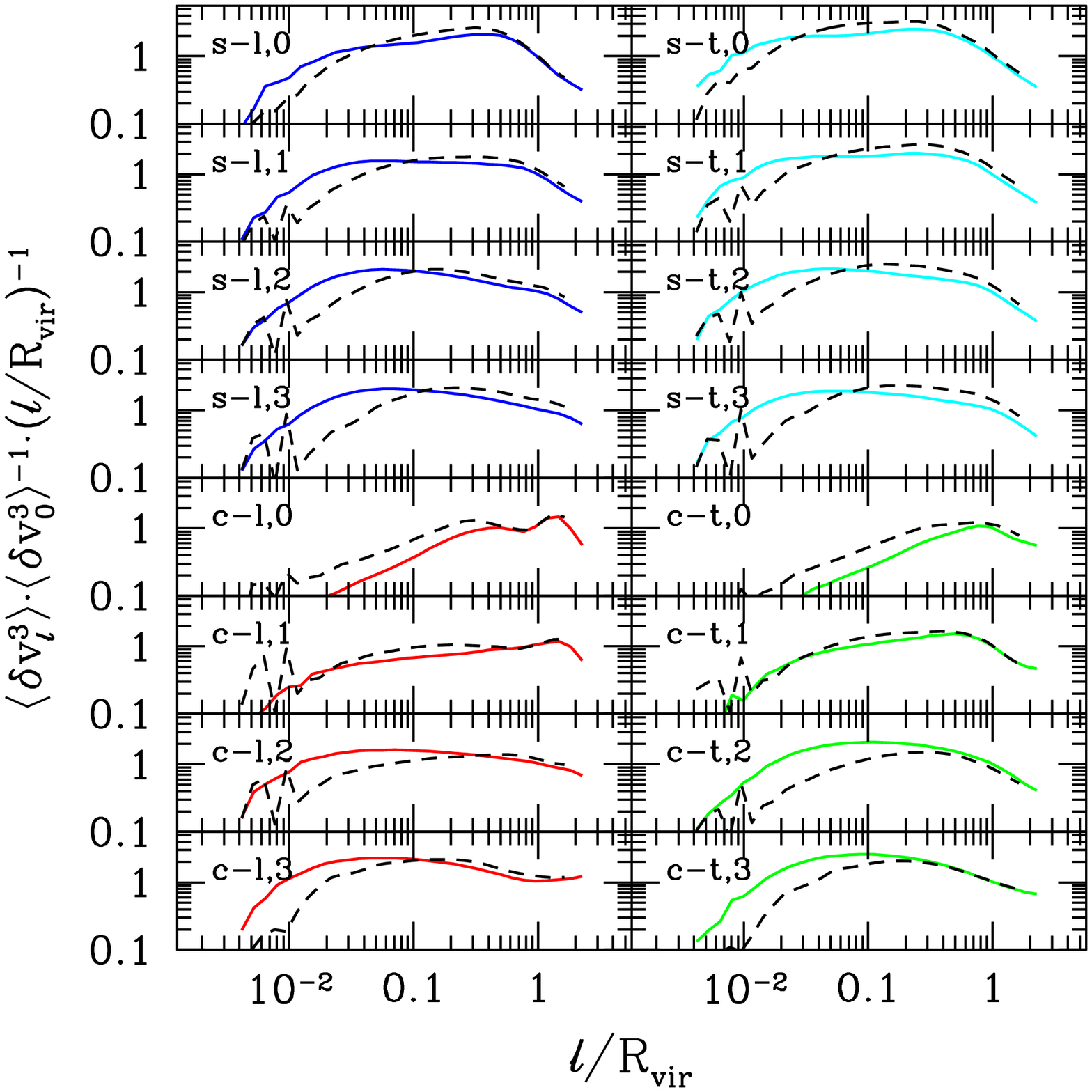}
\caption{ Compensated second (left) and third (right) order structure
  function of the velocity field from our fiducial run (solid line)
  and an adaptive-mesh-refinement run (dashed line).
  The quantity $\delta v_0$ entering the normalization
factor corresponds to the velocity increment on scales $\simeq$\rvir, as measured in the fiducial run.  Labels
  s-l,s-t,c-l,c-t indicate, respectively,
  solenoidal-longitudinal (blue), solenoidal-transverse (cyan),
  compressible-longitudinal (red), and compressible-transverse (green) component.
  Labels 0,1,2,3 correspond, respectively, core
  ($R<(1/3)R_\mathrm{vir}$), off-core ($(1/3)R_\mathrm{vir}<
  R<(2/3)R_\mathrm{vir}$), virial ($(2/3)R_\mathrm{vir}<
  R<R_\mathrm{vir}$), and off-virial
  ($R_\mathrm{vir}<R<(4/3)R_\mathrm{vir}$) region.
  \label{t_f16:fig}}
\end{figure*}
\begin{figure*}[t]
\centering
\includegraphics[width=0.5\textwidth]{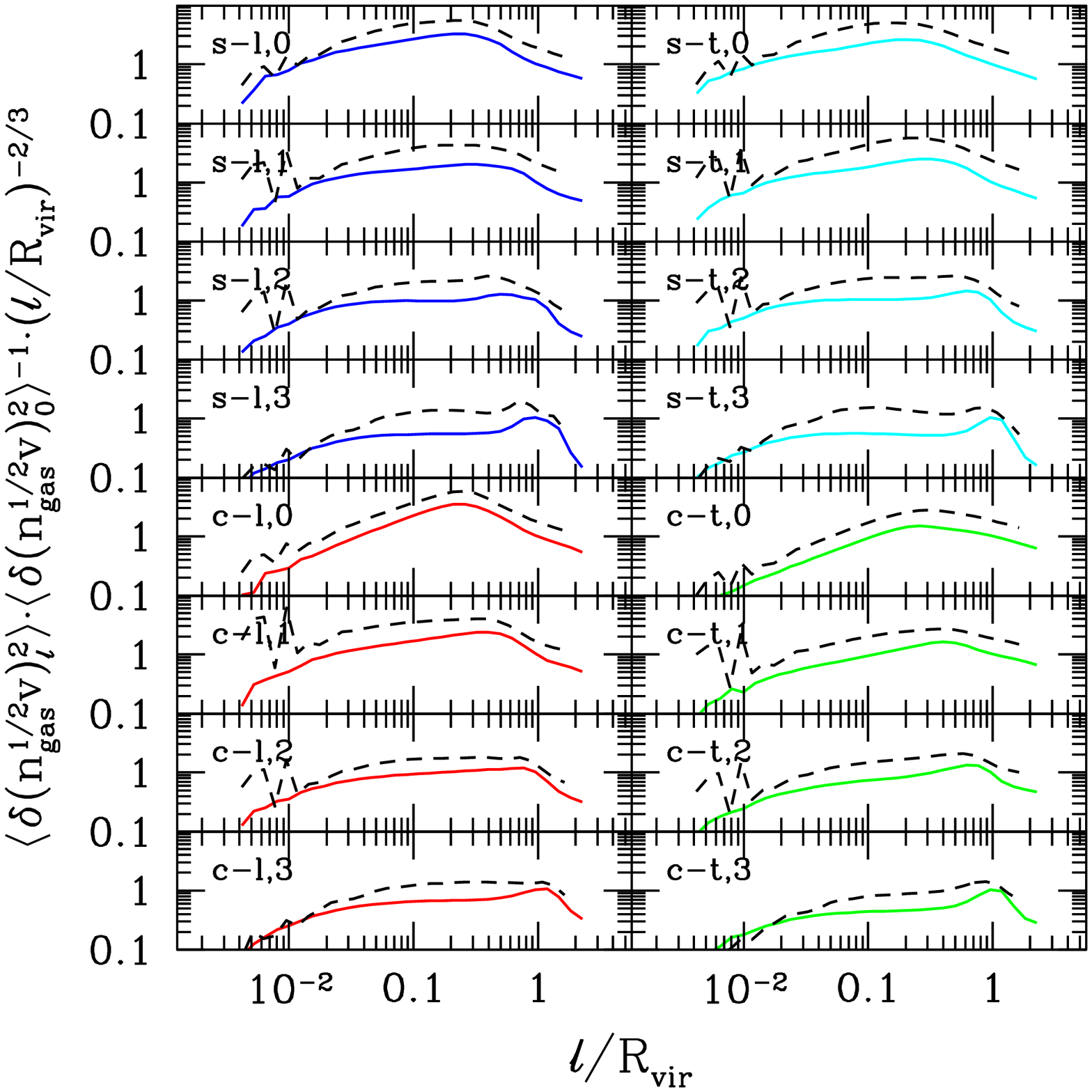}\includegraphics[width=0.5\textwidth]{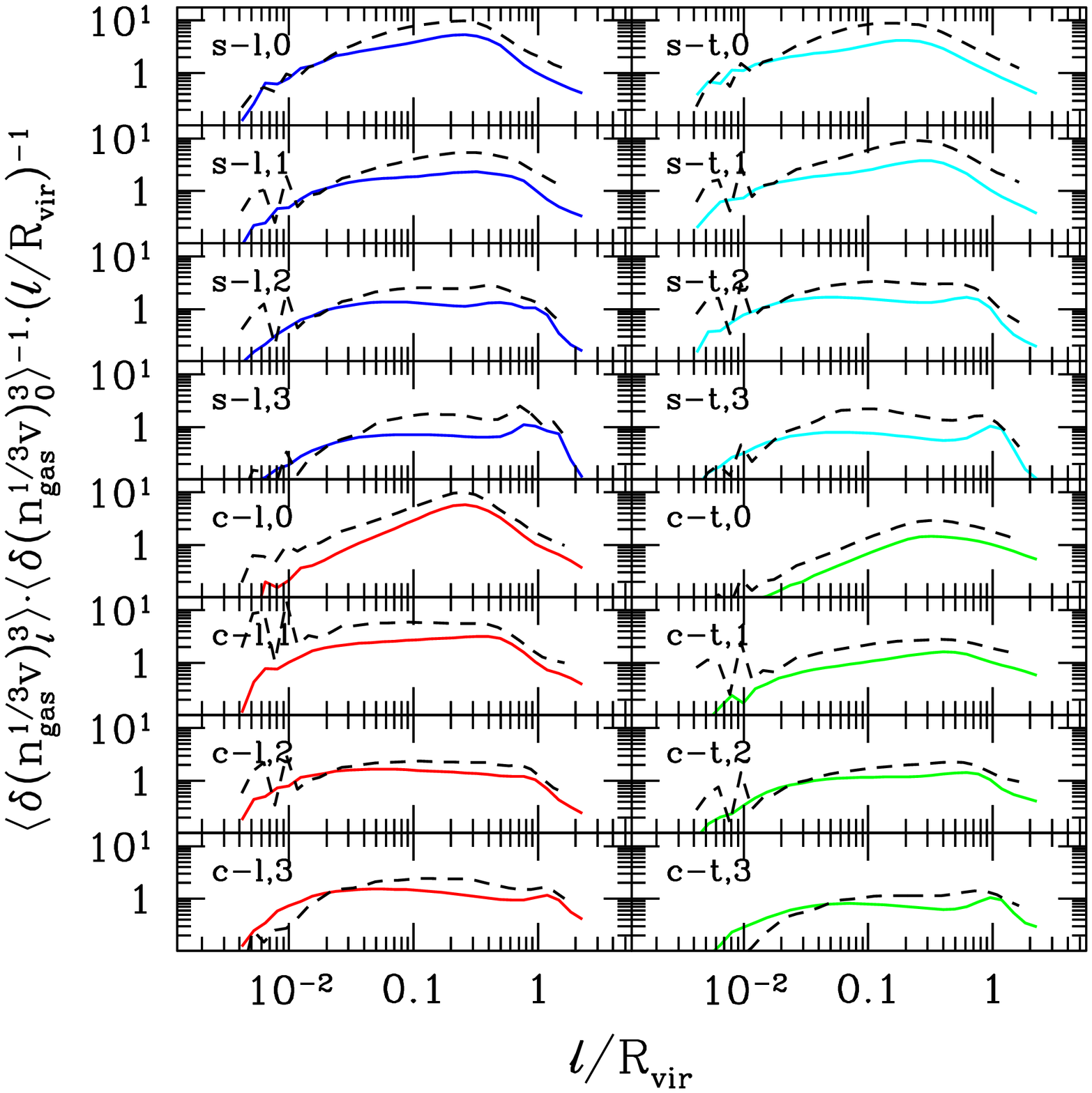}
\caption{Compensated density weighted second (left) and third (right) order structure
  function of the velocity field from our fiducial run (solid line)
  and an adaptive-mesh-refinement run (dashed line).
  The quantity $\delta v_0$ entering the normalization
factor corresponds to the velocity increment on scales $\simeq$\rvir, as measured in the fiducial run.  Labels
  s-l,s-t,c-l,c-t indicate, respectively,
  solenoidal-longitudinal (blue), solenoidal-transverse (cyan),
  compressible-longitudinal (red), and compressible-transverse (green) component.
  Labels 0,1,2,3 correspond, respectively, core
  ($R<(1/3)R_\mathrm{vir}$), off-core ($(1/3)R_\mathrm{vir}<
  R<(2/3)R_\mathrm{vir}$), virial ($(2/3)R_\mathrm{vir}<
  R<R_\mathrm{vir}$), and off-virial
  ($R_\mathrm{vir}<R<(4/3)R_\mathrm{vir}$) region.
  \label{t_f17:fig}}
\end{figure*}

In Section~\ref{tdm:sec} we have briefly addressed the difficulty of a
Lagrangian AMR approach to study statistical properties of turbulence
in the ICM. In particular we have compared a two dimensional density
map from our fiducial run employing Eulerian refinement
(Figure~\ref{t_f3:fig}), with a corresponding map obtained from a
Lagrangian AMR run with the same initial conditions and the same
number of refinement levels (Figure~\ref{t_f4:fig}).  In this section
we provide a more quantitative analysis of the results obtained with
the two approaches, although not an exhaustive one.  In
particular, we will compare second and third order structure function
computed for different velocity components of the velocity increments
and in different volumes of the GC, with and without density
weighting. Results for the ordinary structure functions 
are summarized in Figure~\ref{t_f16:fig},
which is completely analogous to Figure~\ref{t_f15:fig} except that
the dashed line now refers to the Lagrangian AMR run. In
particular, the solid line refer to our fiducial Eulerian AMR run, the
left and right panel correspond to second and third order statistics,
and all labels (and colors) have exactly the same meaning.

As for the solenoidal components (s-l and s-t), there is clearly a
lack of power of kinetic energy on small scales in the Lagrangian AMR
run compared to our fiducial run. The effect becomes more severe
towards the outer regions of the GC, consistent with the qualitative
impression inferred from the comparison of density maps in
Figure~\ref{t_f4:fig}.  In fact, in the outer regions of the GC the
mass density is lower and the Lagrangian refinement is less
effective. The lack of power on small scale is comparable to, although
slightly more pronounced than, in the low resolution run presented in
the previous section. However, on large scales, the structure
functions in the Lagrangian AMR run exhibit excess of power with
respect to both the fiducial and low resolution runs, so they are
qualitatively different. The origin of this feature is not completely
clear. However, we suspect that the bottleneck effect is once again at
work.  Given the relatively low numerical resolution in the outer
parts of the GC, the energy cascade through the nonlinear hydrodynamic
terms is inefficient, so power of kinetic energy accumulates on large
scales. The fact that the issue becomes more visible towards the outer
regions supports this conclusion, although (obviously) a more thorough
analysis is required for a full understanding.

As for the compressional component, (c-l and c-t), its behavior
appears peculiar. Apparently in the core and perhaps even the off-core
regions (0,1), the structure functions in the Lagrangian AMR run are
characterized by excess of power instead of lack thereof.  Towards the
outer regions, the imbalance is reversed and both in the virial and
off-virial region there is a severe lack of kinetic energy on small
scales, as for the solenoidal components. Unlike the solenoidal
components, however, we do not see appreciable excess of kinetic
energy of compressible motions on large scales. This is perhaps again
due to the difference in dissipation mechanism for this component,
which decays in significant part through weak shocks.

Finally, in Figure~\ref{t_f16:fig} we compare density weighted
structure functions computed in Figure~\ref{t_f14:fig}, with those
computed using the Lagrangian AMR run. The comparison shows large
discrepancy between results from the two calculations. In particular,
in the Lagrangian AMR case the density weighted structure functions
always show excess of power at all scales except, occasionally, on the
shortest ones (which are not reliable anyway). The reason for the
spurious result is ascribed to the fact that sampling of the velocity
field in Lagrangian AMR is already biased towards high density
regions. Therefore, when density weighting is applied, the bias
becomes even stronger and superlinear. We conclude that the density
weighted structure functions applied to data based on Lagrangian
adaptivivty are not reliable.

\section{Summary and Conclusions} \label{t_disc:sec}

In this paper we have carried out a numerical study of the turbulence
in the ICM of a massive GC. In order to achieve the necessary dynamic
range of spatial scales across the GC virial volume, we have employed
a novel resolution technique, which we refer to as Eulerian, to
distinguish from the mass threshold based Lagrangian refinement
criterion.  In the Eulerian approach the entire volume occupied by the
GC, or at least a large fraction of it, is progressively refined at
different stages during its gravitational collapse.  This allows us to
study in great detail the statistical properties of the turbulence
that develops inside the GC volume.

We analyzed the mechanism responsible for injecting the turbulence in
the ICM, which we reduce to tidal fields and merging substructures. We
estimate the timescale associated with the largest eddy turnover and
find it of order the GC crossing or dynamical time,
\rvir$/v_\mathrm{vir}$. This time is always $\ll H^{-1}(z)$,
independent of redshift, so turbulent flows should exist inside
collapsed structures independent of redshift.  This is confirmed by
the large values of the vorticity, even on scales comparable to the
virial radius. It is also confirmed by analysis of the structure
functions and in particular the ratio of the transverse to
longitudinal components of the solenoidal second order structure
functions. At least for well resolved scales, this ratio is found in
good agreement with analytic predictions for fully developed isotropic
and homogeneous turbulence.

Analysis of the pdf's of the vor1ticity, velocity divergence and Mach
number consistently indicate that the turbulence is compressible but
only mildly so. Particular features are recognized in the pdf for the
vorticity and velocity divergence that are also seen in dedicated
periodic-box simulations of fully developed compressible turbulence.

Intriguingly, shocks are not the only source of vorticity in the ICM,
not even the dominant one. In fact the baroclinic term is generated by
shocks only in 60\% of the cases in the inner \rvir/3 and in 40\% of
the cases beyond that radial distance. In fact, owing to the complex
assembly history of its constituent substructures, the presence of
internal shocks and lack of complete mixing, the ICM is generally
baroclinic and not barotropic, and vorticity arises even in the
absence of shocks.

Inspection of the structure functions of second and third order in
general indicates that a well defined inertial range of turbulent
cascade is established inside the virial volume and even beyond.  The
injection scale is of order the virial radius, but tends to increase
towards the GC outskirts.  If we apply a Hodge-Helmholtz decomposition
we find that the solenoidal component of the turbulence strongly
dominates in the core region, and while still dominating, becomes
comparable to the compressible component around the virial radius.  In
the core region the structure functions for the solenoidal component
is well described by a Kolmogorov's spectrum while the structure
functions for the compressible component is significantly steeper,
close to Burgers' spectrum.  In the outer regions the structure
functions in general become flatter, indicating perhaps that the
turbulence is injected on multiple scales.

We have also carried out an identical calculation of the same GC using
the same initial conditions, cosmology and finest resolution, but with
the AMR technique based on a mass threshold Lagrangian refinement
criterion.  Compared to our fiducial run, in the Lagrangian AMR case
the solenoidal components of ordinary second and third order velocity
structure functions lack power on small scales while they exhibit
power excess on large scales. These issues become more prominent in
the GC outer regions, where the Lagrangian AMR refined volume
decreases. The compressional component, on the other hand, shows extra
power at all scales in the inner GC regions, and the same severe lack
of power at small scales, in the virial region and beyond.  The
density weighted velocity structure functions extracted from the
Lagrangian AMR run, however, appear strongly affected by bias towards
high density regions. Therefore, such statistics applied to data based
on Lagrangian adaptivity, are not unreliable.

The analysis presented here will be extended to study the statistics
of turbulence at different times during the formation history of the
cluster. This will help us understand how the turbulence evolves
during the cluster formation. The results will be used to understand
how the magnetic field grows and, in particular, at what scales we
should expect equipartition between magnetic and kinetic energy for
given initial conditions as well as to constrain models of
acceleration of relativistic electrons in galaxy clusters.

\acknowledgements 
We are grateful to an anonymous referee useful comments that helped
improve the manuscript.  This work was supported by a grant from the
Swiss National Supercomputing Center (CSCS) under project ID S275.

\bibliographystyle{apj}
\bibliography{../biblio/books,../biblio/codes,../biblio/papers,../biblio/proceed}

\end{document}